\documentclass[fleqn,usenatbib]{mnras}

\usepackage{newtxtext,newtxmath,cuted,lipsum}
\usepackage[T1]{fontenc} \usepackage{graphicx}	\usepackage{amsmath}	\usepackage{amsfonts}
\usepackage{enumerate} \usepackage{colordvi} \usepackage{bm}\usepackage{epsfig}
\usepackage{float}
\usepackage{verbatim}
\usepackage{subfig}
\usepackage[dvipsnames]{xcolor}
\usepackage{multirow,multicol}

\usepackage[colorinlistoftodos]{todonotes}
\usepackage{ulem}
\usepackage{hyperref}

\mathchardef\mhyphen="2D

\newcommand{\hsolar}{{\rm M}_\odot/h}
\newcommand{\hmpc}{{\rm Mpc}/h}
\newcommand{\origami}{{\scshape origami}}

\definecolor{ForestGreen}{rgb}{0.1,0.4,0.1}

\newcommand{\mnedit}[1]{#1}

\title[Intergalactic filaments spin]{Intergalactic filaments spin}  

\author[Xia et al.]{Qianli Xia$^{1}$,
Mark C.\ Neyrinck$^{2,3,4}$,
Yan-Chuan Cai$^{1}$,
Miguel A.\ Arag\'{o}n-Calvo$^{5}$\thanks{E-mails: qx211@roe.ac.uk, Mark.Neyrinck@gmail.com, cai@roe.ac.uk, maragon@astro.unam.mx}
\newauthor
\\
$^{1}$Institute for Astronomy, University of Edinburgh, Royal Observatory, Blackford Hill, Edinburgh, EH9 3HJ, U.K. \\
$^{2}$Ikerbasque, the Basque Foundation for Science\\
$^{3}$Dept of Physics, EHU/UPV University of the Basque Country, Bilbao, Spain \\
$^{4}$Donostia International Physics Center, San Sebasti\'{a}n, Spain \\
$^{5}$Instituto de Astronom\'{i}a, UNAM, Apdo.\ Postal 106, Ensenada 22800, B.C., M\'{e}xico\\
}

\date{Accepted XXX. Received YYY; in original form ZZZ}

\pubyear{2021}

\begin{document}
\label{firstpage}
\pagerange{\pageref{firstpage}--\pageref{lastpage}}
\maketitle

\begin{abstract}
Matter in the Universe is arranged in a cosmic web, with a filament of matter typically connecting each neighbouring galaxy pair, separated by tens of millions of light-years. A quadrupolar pattern of the spin field around filaments is known to influence the spins of galaxies and haloes near them, but it remains unknown whether filaments themselves spin. 
Here, we measure dark-matter velocities around filaments in cosmological simulations, finding that matter generally rotates around them, much faster than around a randomly located axis. It also exhibits some coherence along the filament.
The net rotational component is comparable to, and often dominant over, the known quadrupolar flow. 
The evidence of net rotations revises previous emphasis on a quadrupolar spin field around filaments.
The full picture of rotation in the cosmic web is more complicated and multiscale than a network of spinning filamentary rods, but we argue that filament rotation is substantial enough to be an essential part of the picture. It is likely that the longest coherently rotating objects in the Universe are filaments.
Also, we speculate that this rotation could provide a mechanism to generate or amplify intergalactic magnetic fields in filaments.
\end{abstract}


\section{Introduction}
The cosmic web of bubble-like voids, separated by walls, filaments and haloes, describes the matter distribution of our Universe on scales much larger than galaxies. It is predicted by the standard model of cosmology \citep{klypin/shandarin:1983,bond/kofman/pogosyan:1996} and observed in galaxy surveys \citep{deLapparent/etal:1986,colless/etal:2001,zehavi/etal:2011}. Intergalactic filaments are the skeletons of the cosmic web, generally connecting pairs of neighbouring galaxies.
They feed matter into nodes of the cosmic web, and halo and galaxy spin is known to correlate with orientation with respect to filaments; small haloes' spin vectors tend to align with the axis of filaments they inhabit, while large haloes' vectors tend to be perpendicular to the axis \citep{aragoncalvo/etal:2007,hahn/etal:2007,paz/etal:2008,tempel/libeskind:2013,dubois/etal:2014,codis/pichon/pogosyan:2015,peng2017,ganeshaiahveena/etal:2018,peng2018_1,peng2018_2,kraljic/dave/pichon:2020}. This finding for small haloes is one motivation for our study: small haloes deeply embedded in a spinning filament might tend to spin along with it.

\mnedit{The current understanding of spin in the cosmos begins with the tidal torque theory \citep[TTT,][]{Hoyle1949,peebles:1969}, a formalism that has seen rather continuous conceptual study \citep[e.g.][]{white:1984,PorcianiEtal2002a,PorcianiEtal2002b,Shafer2009,LopezEtal2019,motloch/etal:2020,neyrinck/etal:2020,LopezEtal2021}. In the TTT approximation, a collapsing object is approximated by an aspherical blob of matter with a moment of inertia that generally does not align with its tidal tensor. This misalignment causes the blob to torque up as it collapses, an idea that applies equally well in 2D (i.e.\ to a filament cross-section) as 3D (i.e.\ to a halo, the usual context for these ideas).

The same tidal field that is involved in spin generation can also be used to understand the formation of both filaments, and the haloes that typically form at their endpoints. At these haloes (large density peaks), the tidal field generally prescribes inward flow in all directions. Along density ridges that join these nearby peaks, the tidal field generally prescribes filament formation, i.e.\ expansion along the axis between the peaks, but (nearly cylindrical) collapse along the two directions perpendicular to it \citep{vandeWeygaert/Bertschinger1996,bond/kofman/pogosyan:1996,vandeWeygaert/Bond:2008}. 

At the same time as a filament collapses, the tidal field typically produces a quadrupolar pattern in the vorticity and `spin field' around it \citep{pichon/bernardeau:1999, codis/etal:2012,codis/pichon/pogosyan:2015}. The spin field ($\bm{r} \times \bm{v}$, with $\bm{r}$ being the vector between the filament axis and a particle, and $\bm{v}$ its velocity) around an ideal filament points along the filament axis; it alternates sign across four quadrants,} separated by the primary and secondary axes of collapse. Importantly, the total angular momentum (which we also call `spin'), summing over the four quadrants, is usually \mnedit{ignored}. The spin is precisely zero if integrating over a cylinder in the irrotational initial conditions, but is generally nonzero at later times. The question we ask is whether filaments spin substantially in the late-time Universe.

Filamentary rotation has been speculated about before. There has been work on filaments providing a `swirling, rotating environment' that ends up spinning up nearby haloes \citep{codis/etal:2012,laigle/etal:2015}. In an art-inspired toy `origami approximation,' haloes spin if and only if filaments attached to them do \citep{neyrinck/2016}; this suggests that since galaxies generally spin, so do filaments. Also, haloes spin in 2D cosmological simulations; extruding them to 3D suggests that filaments spin substantially \citep{neyrinck/etal:2020}. But a convincing measurement of how coherently realistic 3D filaments spin is still missing. 

This paper is organized as follows. In \S\ref{sec:methods}, we describe the Millennium Simulation (MS) dataset and methods for our fiducial 3D measurement, using an observationally motivated filament definition. \mnedit{We use the MS because it is a standard dataset in the community, with a good balance between statistics and mass resolution. In \S\ref{sec:results}, we give these measurements. In \S\ref{sec:2d}, as a guide to understanding these 3D results, we show the dynamics in 2D, of `halo' (roughly, filament cross-section) spin in a 2D simulation. This 2D evolution can be tracked much more easily and visually than in 3D, and readers may even wish to read this section before looking in depth at \S\S\ref{sec:methods} and \ref{sec:results}.

In \S \ref{sec:indepth}, we look in further depth, and put our results in the context of a physically motivated filament definition involving (nearly) cylindrical collapse between haloes. For this analysis, we use two further 3D simulations, both with slightly attenuated small-scale modes, which clarifies the dynamics. One of them uses the initial conditions of the Illustris simulation, which will be helpful in our future studies looking at properties of gas and galaxies in filaments. The wide range of simulations used in our study all have similar results, indicating their robustness.}

In \S \ref{sec:object} we give some thoughts on a somewhat subjective issue, of why we consider the `longest coherently rotating objects in the Universe' to be filaments. Finally, in \S \ref{sec:conclusion}, we conclude with a discussion of observational prospects and implications of our results.

\section{Methods}
\label{sec:methods}
For our fiducial measurement, we use a simple filament definition, designed to make straightforward contact with observations: the straight line connecting pairs of dark-matter haloes above a mass threshold.

\subsection{The Millennium simulation}
The Millennium simulation \citep[MS;][]{springel/etal:2005} is a dark-matter-only cosmological $N$-body simulation, using $2160^3$ dark matter particles to represent the matter distribution of a model Universe in a cube of 500\,$\hmpc$ on a side. Each dark-matter particle has mass $8.6\times 10^8 h^{-1}M_\odot$. The simulation was run in a fiducial $\Lambda$CDM model with the following cosmological parameters: $\Omega_m = 0.25, \Omega_b=0.045, h=0.73, \Omega_\Lambda=0.75, n_s=1$ and $\sigma_8=0.9$. For fast computation, we randomly downsampled to one-tenth the original particle density.
 
Dark matter haloes were detected in the simulation with a Friends-of-Friends (FOF) \citep{davis/etal:1985} method, resulting in the same halo sample as in the Millennium database \citep{lemson/virgoconsortium:2006}. We define a halo's position as the location of its most-bound particle, involving a Spherical Overdensity approach. See \citet{Cai/etal:2017} for further details.
We use haloes with mass greater than $10^{13}h^{-1}M_\odot$ to mimic a sample of Luminous Red Galaxies (LRGs) \citep{parejko/etal:2013,zhu/etal:2014}, assuming that each halo hosts one galaxy. We choose this galaxy type because it is well-characterized observationally, the target of many large-scale-structure surveys \citep[e.g.][]{dawson/etal:2013}. We end up with 35,300 haloes in the box. Our fiducial filament catalogue consists of the 33,951 halo pairs separated by 6-10\,$\hmpc$.
This distance range is chosen so that the halo centres are substantially farther apart than their own $\sim 1$\,$\hmpc$ radii, but are near enough to be typically connected by a coherent dark-matter filament \citep{colberg/krughoff/connolly:2005}.
For longer separations, there are many more pairs (3,732,470 pairs for the longest separation we considered). To keep the noise level similar for all separations, we analyse 30,000 randomly chosen pairs in each of the longer cases; adding more filaments changed results negligibly.

What does a `coherent dark-matter filament' mean? There is not a universally agreed-upon, absolute definition of this \citep{libeskind/etal:2018}, but our conception, shared by many, is a density ridge \citep[e.g.][]{sousbie:2011} between two haloes. Density ridges are thought to exist even without haloes at both endpoints, but these would be difficult to detect observationally \mnedit{for modest-size filaments, so we only consider filaments with two halo endpoints}. Generally, within high-enough density ridges, dark-matter streams have crossed, i.e.\ initially distant dark matter particles have come together in the same place, moving at different velocities, possible for such a collisionless fluid \citep[e.g.][]{falck/neyrinck/szalay:2012}. Also, that gas has shocked (the baryonic version of stream crossing). We expect that a typical filament in our fiducial catalogue (defined as straight-line segments between LRGs in this length range) corresponds approximately to some density-ridge filament. 
\mnedit{This sort of definition has been used extensively in observations to detect observational signals related to filaments, such as from lensing, and the thermal Sunyaev-Zeldovich effect \citep{clampitt/etal:2016,epps/hudson:2017,deGraaff/etal:2019,xia/etal:2020,tanimura/etal:2019}. This definition is particularly useful when there may be no other reliable tracers of the structure between haloes available.} For comparison, we also test longer filaments in the following ranges: 10-20, 20-40, and 40-60\, $\hmpc$.

\subsection{Filament definition and stacking}
\label{sec:stacking}
\begin{figure}
    \centering
    \includegraphics[width=\columnwidth]{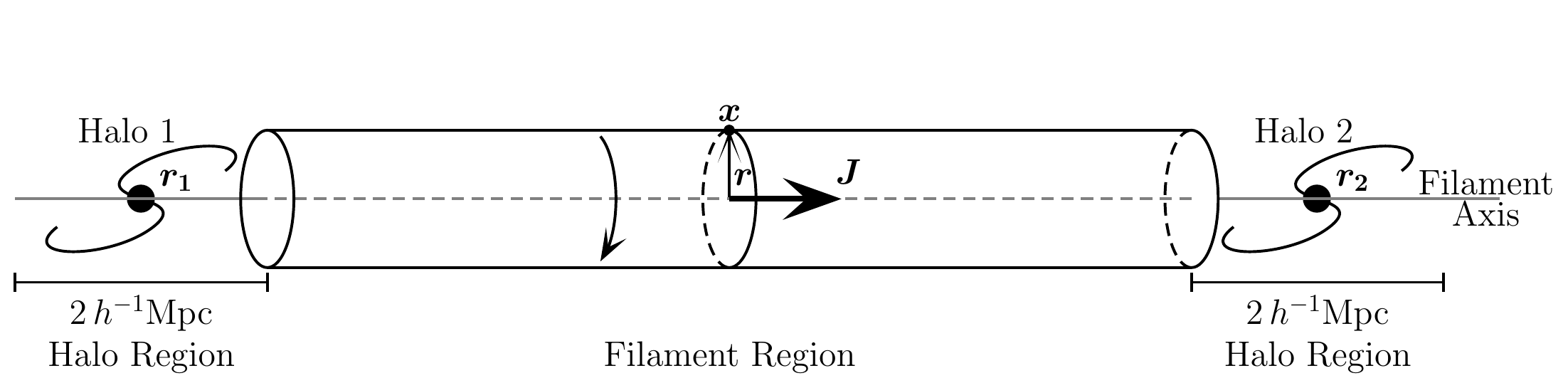}
    \caption{Illustration of coordinates for the halo-filament system. The line connecting the two haloes at $\bm{r_1}$ and $\bm{r_2}$ is defined as the filament axis. The filament region is defined to be 1 $\hmpc$ away from each halo, to isolate the filament signal from possible contamination from flows within each endpoint halo. The average angular momentum $\bm J_{\rm avg}$ (see Eq.~\ref{eqn:javg}) for an individual filament defines the direction of the spin.}
    \label{fig:illustration}
\end{figure}
As shown in Fig.~\ref{fig:illustration}, for each pair of haloes at positions $\bm{r_1}$ and $\bm{r_2}$, we define the filament axis as the line connecting them, with the direction $\hat{\bm{n}} = \frac{\bm{r_2}-\bm{r_1}}{|\bm{r_2}-\bm{r_1}|}$. 
For every particle with position vector $\bm{x}$, $L = (\bm{x}-\bm{r_1}) \cdot \hat{\bm{n}}$ is then the projected distance from $\bm{r_1}$. 

We define the filament region to be the dark matter between the two haloes, and at least 1\,$\hmpc$ away from each halo along the filament axis to avoid confusion from velocites within each halo. The average virial radius $r_{200}$ of our halo sample is 0.5 $\hmpc$, $\sim 1$\,$\hmpc$ in the 99th percentile. Halo regions are the matter within 1\,$\hmpc$ from their centres along the filament axis.
For each particle in the filament region, we calculate the projected angular momentum $J_{\rm proj} = (\bm{r} \times \bm{v}) \cdot \hat{\bm{n}}$.
We divide particles into 15 equally spaced cylindrical shells around the filament axis, ranging from 0 to 6\,$\hmpc$.
Here, we choose the spin direction $\lambda = \pm 1$ for a filament so that the average angular momentum over the cylindrical bins is positive. That is, that
\begin{align}
J_{\rm avg}=\frac{\lambda\sum_{\text{shells }i} \left(\sum_{\text{particles }k} J_{{\rm proj},k;i}\right) \sigma_i^{-2} N_{i}^{-1}}{\sum_{\text{shells }i} \sigma_i^{-2}} > 0, 
\label{eqn:javg}
\end{align}
where $N_{i}$ is the number of particles in the $i$th cylindrical shell, $\sigma_i^2$ is the variance of $J$ over particles in the shell, and the subscript $k$ labels each particle within the shell. To stabilise $J_{\rm avg}$ statistically, we inverse-variance-weight the sum over shells.

\begin{figure*}[H]
	\centering
    \includegraphics[width=0.97\textwidth]{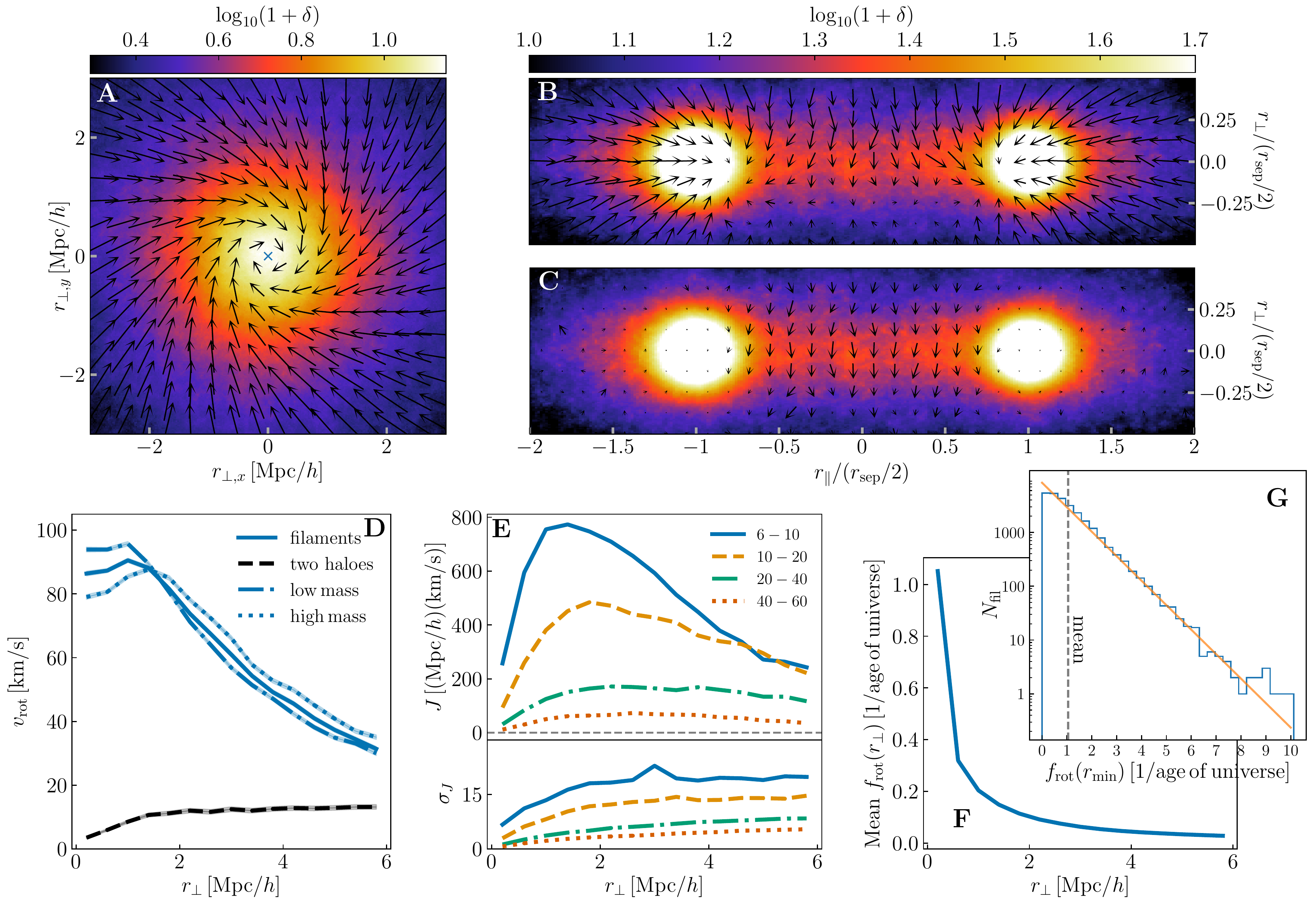}
    \caption{{\em Top panels}: {\bf Average density and momentum density fields around filaments.} Colors show the average density around filaments connecting pairs of haloes separated by $6$-$10$\,$\hmpc$ found in the MS. Fig.~\ref{fig:illustration} is a sketch of the coordinate system.
    ({\bf A}) Arrows show the momentum density $\bm{p}$ in the filament region, projected along the filament axis, excluding the halo regions. The longest vector corresponds to 1380 km/s.
    ({\bf B}) Lengthwise view on one side of the filament axis, stacked over filaments (rescaling coordinates so each halo lies at $\pm 1$), and projected. Here, arrows show $\bm{v}$ instead of $\bm{p}$, for clarity at large $1+\delta$. The longest vector corresponds to 264 km/s. ({\bf C}) Same as (B), but with infall velocities around two haloes nulled (See text). 
    {\em Bottom panels}: ({\bf D}) {\bf Rotational velocity profile, and mass dependence.} Average azimuthal velocity $v_{\rm rot}$ as a function of distance to the filament axis $r_\perp$, in the filament region and the halo regions as shown in panel B. Matter in the filament region (blue) exhibits stronger rotation around the filament axis than in halo regions (black). \mnedit{Dotted and dashdotted blue curves show the profiles for low- and high-mass subsamples, defined to contain filaments in the lowest and highest 30\% in the mean mass of the two haloes.} ({\bf E}): {\bf Filament angular momentum density profile, $J(r_\perp)$.} The upper panel compares $J(r_\perp)$ for different filament lengths in units of $\hmpc$. Signals from random patches of the same length in the simulation have been subtracted for each case. The lower panel shows their corresponding error bars, which are the standard errors of the mean velocity of a filament in the filament stack. ({\bf F}) {\bf Rotation frequency.} The mean frequency at each distance, excluding any filaments that counter-rotate (compared to $J_{\rm avg}$) in that distance bin.  ({\bf G}): {\bf Rotation frequency distribution} over filaments, in their innermost bins (blue). Excluding the 23\% of filaments that counter-rotate (compared to $J_{\rm avg}$) in this bin, the mean frequency is 1.05, i.e.\ a rotation period of about the age of the Universe. An exponential distribution (orange) with this mean fits the distribution well.} 
    \label{fig:isotropic}
\end{figure*}

Having defined coordinates, we measure filaments' density and angular-momentum density fields in rectangular boxes along the axis, and then project these boxes either along or perpendicular to the axis. \mnedit{To cleanly combine results from different filaments, for each filament we rescale the distance along the filament $r_\parallel$, so that the two haloes are centred at $\pm 1$, but keep the filament/halo region boundary fixed at 1 $\hmpc$ (Fig.~\ref{fig:illustration}).}

\subsection{Measuring rotational velocity and angular momentum density}
We measure the rotational velocity of dark matter in cylindrical shells for the filament and halo regions. 
The average rotational velocity in the $i$th shell is defined as 
\begin{align}
\left<v_{\rm rot}\right>_{\text{shell }i} = \frac{\sum_{\text{filaments }f} \sum_{\text{particles }k} J_{{\rm proj},k;i;f} \cdot \lambda}
{\sum_{\text{filaments }f}  N_{i;f} r_{\perp,i}},
\end{align}
where $N_{i,f}$ is the number of particles within the $i$th shell in the $f$th filament, and $r_{\perp,i}$ is the perpendicular distance to the $i$th shell.
For the average angular velocity in haloes, we repeat above calculation but for particles that are within 1\,$\hmpc$ to the haloes.

\mnedit{We use the term {\it angular momentum density} for a quantity conceptually defined as $J=(1+\delta)v_{\rm rot} r_\perp$, i.e., a quantity that volume-integrates up to the total angular momentum of the filament (divided by the mean density). Note that this is not the same as the curl of the momentum-density field $(1+\delta)\bm{v}$, which is more directly related to the vorticity. Explicitly, we measure $J$ in the $i$th shell as}
\begin{align}
\left<J\right>_{\text{shell }i} = \frac{\sum_{\text{filaments }f} \sum_{\text{particles }k} J_{{\rm proj},k;i;f} \cdot \lambda}
{\sum_{\text{filaments }f}  R_{{\rm sep};f} A_{i} \bar{\rho}},
\end{align}
where $R_{{\rm sep};f}$ is the length of the filament region, $A_{i}$ is the area of $i$th annulus, and $\bar{\rho}$ is the mean density in the simulation.

\section{Results}
\label{sec:results}
\begin{figure*}
	\centering
	\includegraphics[width=\textwidth]{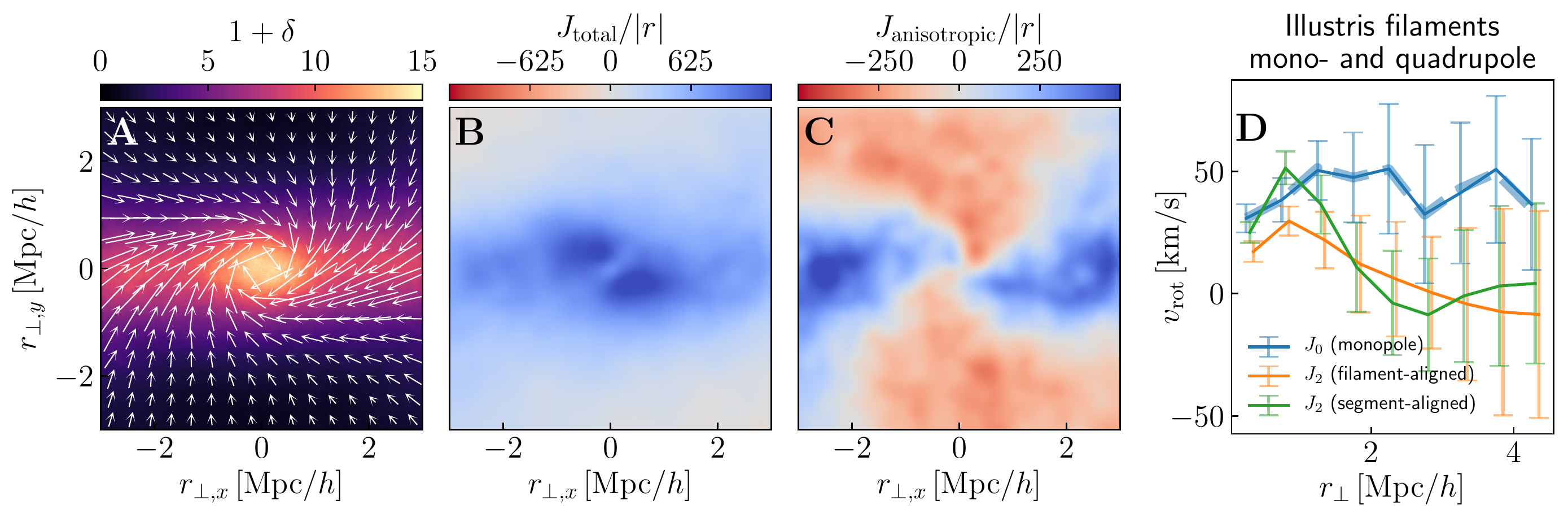}
	
	\caption{{\em Panel A}: Average density (colors) and momentum density fields (vectors) around MS (panels A-C) filaments, as in Fig.~\ref{fig:isotropic}A, but stacked so that the wall in which a filament might be embedded will appear horizontally. The maximum length of vectors corresponds to 1760 (km/s). {\em Panel B}: Magnitude of angular velocity with the positive direction defined as into the paper. {\em Panel C}: Angular velocity with the isotropic component subtracted. The isotropic, monopole component dominates the quadrupole. {\em Panel D}: Monopole ($J_0$) and quadrupole ($J_2$) amplitudes as a function of distance, averaged over 32 filaments of length 6-10 $\hmpc$ from a rerun of the Illustris simulation. These followed the collapsed ridge of the filament, rather than the straight line between endpoint haloes. For the orange `filament-aligned' curve, $J_2$ was measured for each filament; for the green `segment-aligned' curve, $J_2$ was measured separately in each of 6 segments along each filament, resulting in increased $J_2$ amplitude in the inner bins. See  \S\ref{sec:indepth} and Fig.\ \ref{fig:vel_multipole} for further discussion.
	}
	\label{fig:pca}
\end{figure*}

Fig.~\ref{fig:isotropic} shows the projected matter and momentum density $\bm{p}=(1+\delta)\bm{v}$, averaged over 33,951 filaments 6-10\,$\hmpc$ long. ($1+\delta=\rho/\bar\rho$ is the matter density $\rho$ relative to the mean matter density $\bar{\rho}$, and $\bm{v}$ is the velocity.)
In addition to the radial infall due to gravity, there is a clear rotational component around the filament axis.
These two sum up to the spiral pattern shown in Panel A. \mnedit{In the next two lengthwise panels, we rescale the coordinate along each filament axis to put all endpoints at $\pm 1$.} (B) shows that, on average, the pattern is coherent across the filament. 
In (C), for each halo pair with $\lambda = -1$, we flipped the sign of velocities instead of interchanging positions of the pair, before stacking them. This procedure aligns the sense of rotation but flips any infall velocity near the two haloes. When averaged over a large sample, this has effect of nearly nulling flows along the filament.
The rotational velocities close to the two haloes are small; it seems that for this sample of rather large haloes, rotations of their host dark matter haloes are not particularly aligned on average with that of the filaments.
In the filament region (between the haloes, but with distance along the filament $>$1\, $\hmpc$ from both -- see Fig.~\ref{fig:illustration}), the average rotational velocity peaks at around 80 km/s near the filament axis. In (D), the rotation pattern has a peak at 1$\hmpc$, and thus is neither like a solid body (with velocity increasing with distance), nor like a fluid vortex (with decreasing velocity). Also of note is that, in the halo region (outside the filament region), the velocity it is much smaller. Thus the endpoint haloes only slightly corotate with the filament.
We find that for the innermost bin, the period of rotation is of order the age of the Universe, befitting the Universe's longest spinning objects. This Hubble timescale also accords with the magnitude of vorticity generated in caustics \citep{pichon/bernardeau:1999}. The rotation frequency has an exponential distribution, going up to ten times the mean in our sample. Still, this timescale is long compared to the crossing time  of a typical filament; it seems that just like in haloes, radial motions dominate rotational motions for the dark matter, i.e.\ dark-matter filaments are not typically rotation-supported.

\mnedit{We repeated the measurement for filaments with the least and most massive 30\% of halo pairs (judging by the average mass of the pair). Curiously, within $r_{\perp}\sim 1$ $\hmpc$, the rotation is highest for low-mass pairs, but this relationship flips outside this radius. It is tempting to think of this inner regime as `within the filament,' but typical filaments could be curved on scales up to $\sim 1$ $\hmpc$, so we caution against this conclusion.}

To test if the filament-rotation signal might arise from random velocity flows, we repeated the measurement around pairs of random positions with the same separations as the haloes.
The average angular momentum density profile $J(r_{\perp})$ in the random filaments is positive, to be expected since it is on average aligned with $J_{\rm avg}$ in the same way as in the real sample, but in the randoms it is much smaller. Panel E of Fig.~\ref{fig:isotropic} shows that ${J}(r_{\perp})$ from filaments is significantly higher than zero, after subtracting the signal from random filaments. For a plot showing both reals and randoms without the subtraction, see Fig.~\ref{fig:wdm_tags}, using a warm dark matter simulation, with similar results as in the MS. In summary, angular momenta of real filaments are much larger than those of randoms. 

We also investigated how the signal varies with filament length (Panel E). We find that the rotation decreases for longer filaments, becoming very small at 40-60\,$\hmpc$. This is expected, since pairs of haloes at this large separation rarely have nearly straight-line physical density ridges connecting them.

\subsection{Relation to quadrupolar flows}
\label{sec:model}
\begin{figure*}[H]
    \includegraphics[width=\textwidth]{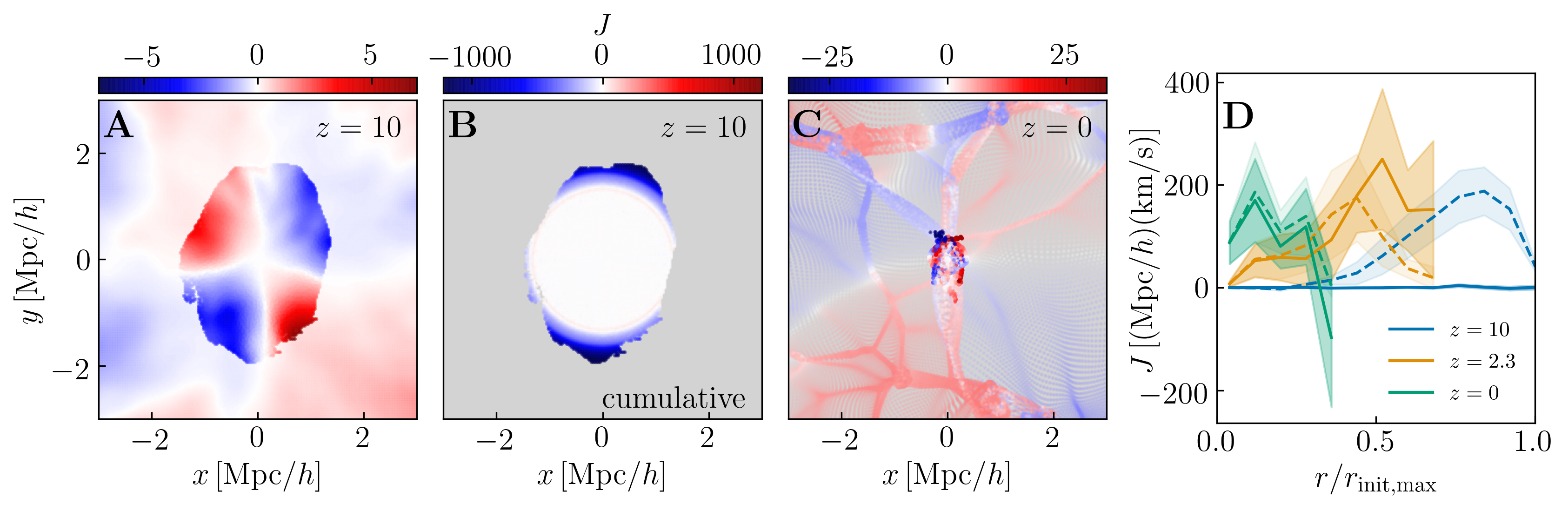}
    \caption{\textbf{Evolution of halo angular momentum in a 2D simulation.} If we neglect variation along a 3D filament, this halo from a 2D simulation corresponds to a filament cross-section. Colorbars in Panels A-C apply to the angular momentum $J$. In Panels A \& C, particles in the simulation that collapse to this halo by $z=0$ are shown with bold colours; other particles are shown with muted colours.  (\textbf{A}) At $z=10$, the quadrupole pattern is clear both for the protohalo region and the surroundings. (\textbf{B}) The cumulative net angular momentum of particles, summed from the centre, shows that $J$ originates at the outskirts of the region that later collapses. (\textbf{C}) The quadrupole randomizes during evolution from $z=10$ to the current epoch (for clarity, $J$ for the non-collapsed particles has been transformed to $\sinh^{-1}(J/10)/2$.)
    (\textbf{D}) The average angular momentum in radial bins from 183 such 2D haloes at three different epochs shown in the legend. We see the angular momentum being transported to the centre. Here $r_{\rm init, max}$ is the largest comoving distance from the area centroid to a particle in the initial conditions, and $r$ is in comoving coordinates. Solid curves include all particles, while dashed curves include only particles that are collapsed at $z=0$. The shaded regions represent the error on the mean. The amplitude is lower than in Fig.\ \ref{fig:isotropic}E; note though the visible noise from the smaller sample here.}
    \label{fig:2d_sim}
\end{figure*}

This coherent filament rotation has not before been measured, to our knowledge. How is it related to the quadrupolar spin field mentioned above \citep{codis/pichon/pogosyan:2015}? This finding is also related to the quadrupolar vorticity field expected around filaments at late times \citep{pichon/bernardeau:1999,laigle/etal:2015}. (Recall, however, that we are showing the angular momentum density, rather than vorticity here.) It arises from the curl-free, gravitationally-sourced velocities of an elliptical density peak (\S\ref{sec:model}); also see Fig.\ \ref{fig:2d_sim}.
The orientation of each filament cross-section is random in Fig.\ \ref{fig:isotropic}, averaging out any such pattern, so in Fig.~\ref{fig:pca} we orient the stack along the major axis (the largest eigenvector) of the projected matter distribution around each filament. 
The density concentration along the $x$-axis indicates a wall on average, within which the filament is embedded (left panel in Fig.~\ref{fig:pca}). The overall spin is positive everywhere, with its amplitude varying with orientations (middle panel).

Subtracting the isotropic signal (measured from the left panel of Fig~\ref{fig:isotropic}) reveals a quadrupole pattern (right panel in Fig~\ref{fig:pca}). Importantly, the average amplitude of the monopole, which was not considered in the past, exceeds or is comparable to the average quadrupole. This suggests that it should not be ignored when interpreting the alignment of galaxy spins with their filaments. The white zero regions separating the quadrupole regions grow vertical and horizontal at large radius, as expected \citep{codis/pichon/pogosyan:2015}. But closer to the filament, the quadrupole is $\sim45^\circ$ rotated, perhaps from the matter flow overshooting the filament with some impact parameter.

\begin{figure}
    \centering
    \includegraphics[width=\columnwidth]{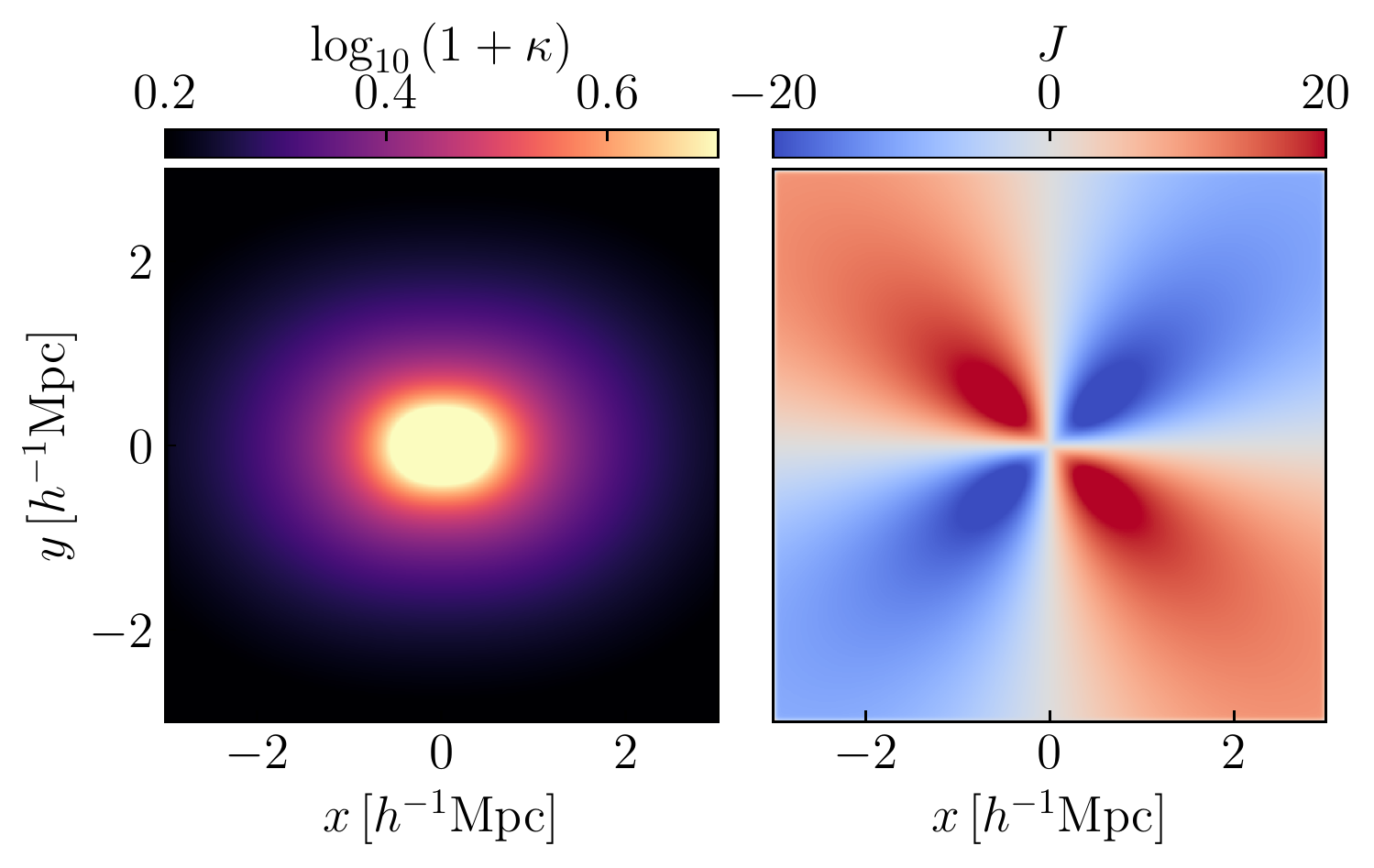}
    \caption{An idealized elliptical density peak (Left). The gradients of its gravitational potential give rise to a quadrupolar pattern of angular momentum density (Right).}
    \label{fig:model}
\end{figure}

To illustrate a simple situation in which a quadrupolar pattern of angular momentum arises, we show a 2D elliptical density peak in Fig.~\ref{fig:model}. Following Eq.\ (12) in \citet{mandelbaum/etal:2006}:
\begin{align}
\Phi(r, \theta)=\frac{2 A}{(\alpha-2)^{2}} r^{2-\alpha}\left[1+\frac{B(\alpha-2)^{2}}{\alpha(\alpha-4)} \cos (2 \theta)\right],
\end{align}
with parameters $\alpha=0.8,e_h=0.3,A=1000$ and $B = e_h \alpha /2$ for illustration. The convergence $\kappa$ is then computed via
\begin{align}
    \kappa=\frac{1}{2} \nabla^{2} \Phi
\end{align} using Fourier transforms.  

We use the Zel'dovich approximation \citep{zeldovich:1970} to estimate the gravitationally-sourced velocity from this, taking the gradient of the inverse Laplacian of the 2D density field, i.e., $\bm{v} = \nabla \Phi$. Note that these velocities are exact in the initial conditions, and should be good approximations on large scales at late times.
Finally, we compute the angular momentum density using the modelled density and velocities.
The result is shown in the right panel of Fig.~\ref{fig:model}; a pure quadrupole with no monopole. A typical quadrupole arises simply from the potential flow that an elliptical density peak produces. This has no monopole in the initial conditions, and the monopole that we see at later times comes from fluctuations away from this in halo shape, in the density field or in its Lagrangian boundary.

\subsection{Rotation signal from random samples}
\label{sec:randoms}
Our measure of angular momentum is by construction positive, so its positive average is meaningless. To attach physical meaning, it is essential to know the amplitudes of the signal arising from a random scenario. For this, we define pairs of random points chosen to have the same lengths as the real filament samples. We repeat measurements as shown in the previous sub-section for the angular momentum density profiles for these random samples. For most of the cases we have investigated, the random signal is found to be sub-dominant over the signal from real filaments. The profiles shown in Fig.\ \ref{fig:isotropic}\mnedit{E} have had the random signal subtracted off. For the longest filaments (40-60\,$\hmpc$), the residual $J(r_\perp)$ is very small, indicating that the signal is very close to the random case. See Fig.\ \ref{fig:wdm_tags} for a comparison of typical real and random signals, without subtracting off.  In estimating the error bars shown, we neglected the small error on $J(r_\perp)$ measured from the randoms.

\begin{figure}
    \centering
    \includegraphics[width=0.6\columnwidth]{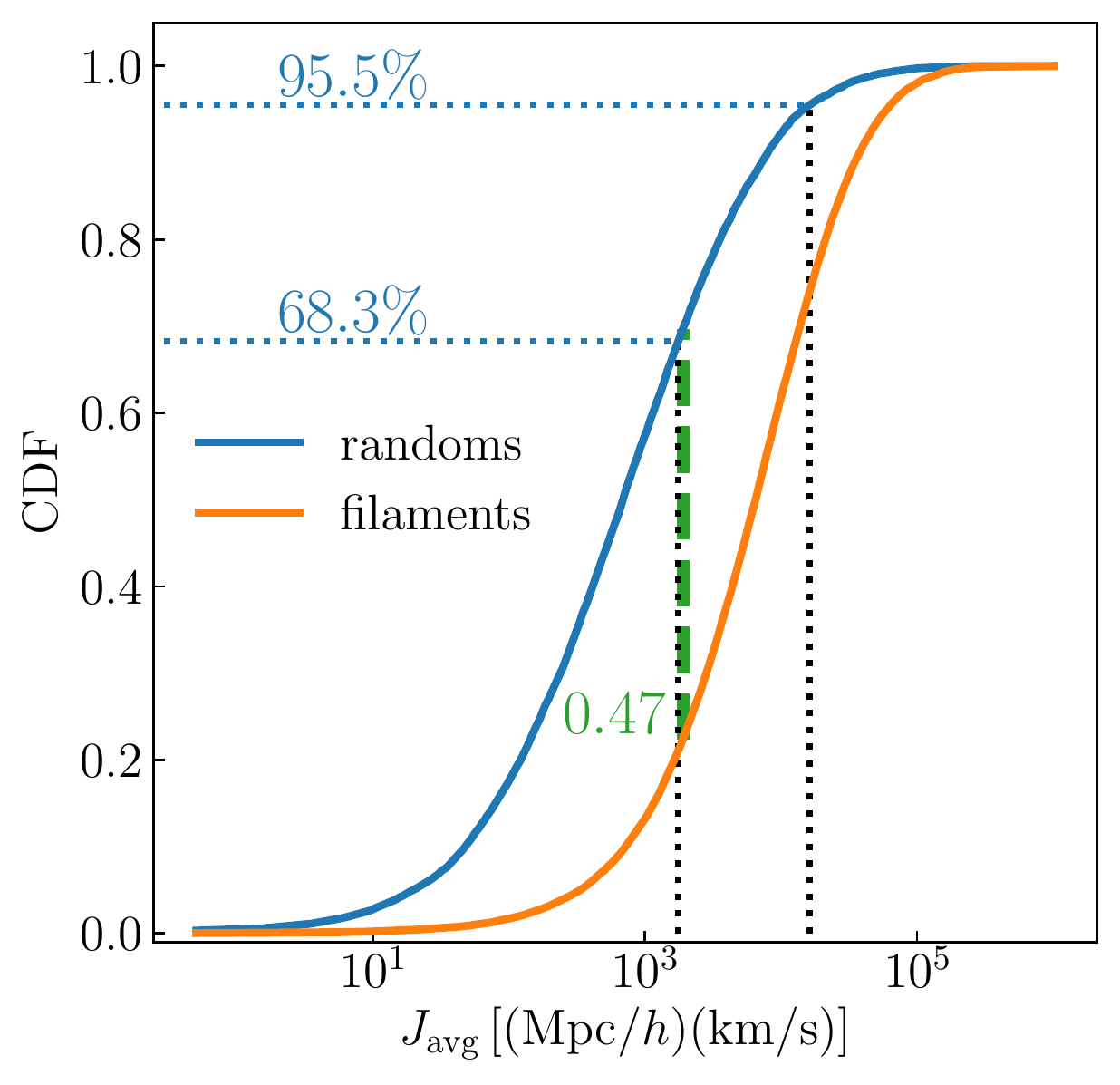}
    \caption{The cumulative distribution functions (CDF's) of the average angular momentum $J_{\rm avg}$ for random line segments and filaments. The green dashed line indicates the maximum difference between the CDF's, the $D$ statistic used in a two-sample Kolmogorov-Smirnoff test. The black dotted lines indicate the thresholds corresponding to 68.3\% (1-$\sigma$) and 95.5\% (2-$\sigma$) probabilities that $J_{\rm avg}$ is not drawn from the $J_{\rm avg}$ distribution of random line segments. In the random sample, the fractions exceeding these thresholds are $1-0.683=31.7$\% and $1-0.955=4.5$\%, to be compared respectively with 79\% and 26\% of real filaments.}
    \label{fig:CDF}
\end{figure}

\begin{figure}
    \centering
    \includegraphics[width=0.6\columnwidth]{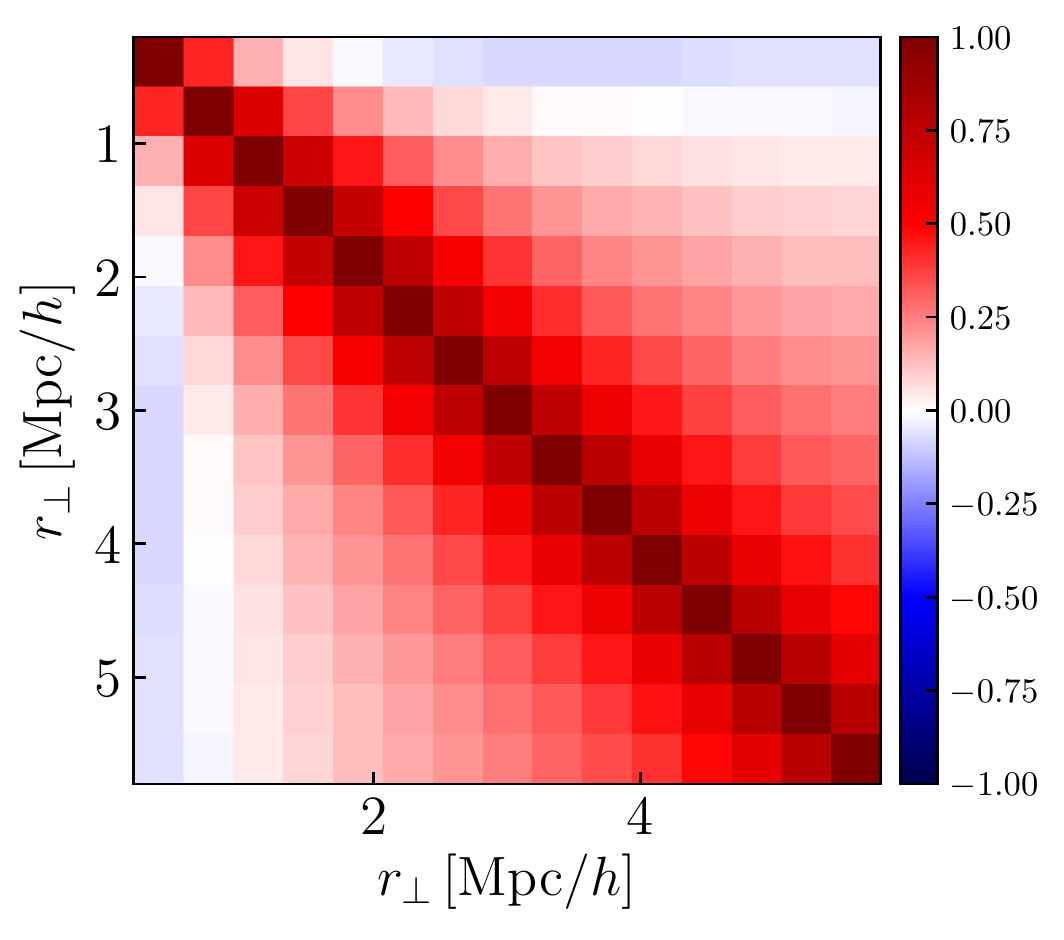}
    \caption{Filament-to-filament correlation matrix of the angular momentum density $J(r_{\perp})$ in different radial bins. Each sample is a single filament with length 6-10\,$\hmpc$. The positive off-diagonal elements indicate that matter in different radial bins  indeed generally co-rotate. This correlation decreases with increasing bin-separation, as expected.}
    \label{fig:corr}
\end{figure}

A standard, single summary statistic of the difference between such distributions uses the two-sample Kolmogorov-Smirnov test. We find a maximum vertical separation between the CDFs of $D = 0.47$. For this sample of tens of thousands, the null hypothesis that the two samples are drawn from the same distribution is rejected with essentially certain confidence; the $p$-value characterizing the probability that the two distributions are the same is $\sim 10^{-120}$.

More tangibly, how might we answer the question of whether an individual filament is substantially rotating or not? We can decide by reference to the distribution of $J_{\rm avg}$ for the random sample, shown in Fig.\ \ref{fig:CDF}. If we define a filament to be `rotating' if its $J_{\rm avg}$ is over a 1-$\sigma$ cut (where 31.7\% of the randoms have higher $J_{\rm avg}$), 79\% of the real filaments are rotating. If we are more strict, and define `rotating' with a 2-$\sigma$ cut (where 4.55\% of the randoms have higher $J_{\rm avg}$), 26\% of the real filaments are rotating. This 2-$\sigma$, 26\% fraction is our fiducial, simple response to the question `how often do filaments substantially rotate?' but it uses an admittedly arbitrary threshold.

\subsection{Rotation coherence with radius}
The direction of angular momentum for each filament is defined with an inverse-variance-weighting scheme as described above. The question remains how coherent the rotation is with radius. To check this, we compute the correlation matrix for the rotational velocities $\bm r \times \bm v$ at different radii. An example for the 33,951 filaments with 6-10\,$\hmpc$ separation is shown in Fig.~\ref{fig:corr}. We find that the correlation coefficients between adjacent radial bins are close to unity, indicating that the matter in adjacent cylindrical shells strongly tends to co-rotate. The coefficient decays with bin separation, as expected.

\section{2D simulation}
\label{sec:2d}
In this section, we examine halo collapse and rotation in 2D, as a conceptual guide to filament collapse and rotation in 3D. This correspondence is obviously only approximate. Dynamics in 2D and 3D can qualitatively differ in many physical systems. But the relatively simple dynamics of the cosmic web on megaparsec and larger scales encourages us to think that looking at a simplified 2D setting here can help, as long as we keep in mind the caveats that we are neglecting variation along the filament, and motions out of the plane (such as filament bending, or rotation perpendicular to the axis).

The monopole we find is not present in the primordial, irrotational velocity field of the simulation. Integrating a zero-curl field around a circle, the angular momentum vanishes. Where, then, does the rotation at a later time come from? It comes from the non-circular outskirts of a filament cross-section in the initial conditions, generally carrying angular momentum \citep{Doroshkevich:1970}. To understand the collapse of matter onto a cross-section of a filament, we additionally investigate a two-dimensional $N$-body simulation.

This simulation was previously used by \citet{neyrinck/etal:2020}, with some analysis and animations (see \url{https://youtu.be/7KjesL_hP7c}) relevant to the present work. The simulation has 1024$^2$ particles, and a box size of 32\,$\hmpc$. The initial conditions were a 2D slice of particles from a $1024^3$ simulation with a BBKS \citep{BBKS} initial power spectrum, with modes of wavelength $< 1\, \hmpc$ suppressed in amplitude, to simplify the physical problem and clarify structures like filaments. The ($x$-$y$) plane of particles was replicated along the $z$-axis to match the number of slices in the original particle grid.  It was run using a version of {\scshape Gadget2} \citep{Gadget2}, modified to calculate forces and update positions only in the $x$ and $y$ directions. Because of this replication along the $z$-axis, each particle represents a cylinder, interacting with other cylinders effectively with a 2D version of gravity. So, each of its haloes is literally an infinite filament. For the initial power spectrum and expansion history, we used a generic $\Lambda$CDM ($\Omega_{\rm M}=0.3$, $\Omega_\Lambda=0.7$, $\sigma_8=0.8$, $h=0.7$) set of cosmological parameters.

We detect haloes in the 2D simulation using the \origami\ \citep{falck/neyrinck/szalay:2012} algorithm. A particle is classified as a halo particle if, going from the initial to final conditions, it has `folded' (crossed some other particle) along two (in 2D) initial orthogonal axes. We then join together groups of particles adjacent on the initial Lagrangian square grid to form haloes.\footnote{Several haloes returned in this process were actually groups of haloes apparently distinct by eye, joined by small bridges of halo particles. The spurious fragmentation of filaments into tiny haloes in simulations with truncated initial power \citep{wang/white:2007} likely contributed to this. Applying a mathematical morphology erosion operator cut many of these bridges, returning haloes that generally corresponded to visual expectation. Erosion, used e.g.\ by  \citet{platen/etal:2007}, shaves off cells within a specified distance (here, one pixel) of a boundary from all contiguous blobs. We computed the erosion by smoothing the Lagrangian `halo' (1) and `not-halo' (0) field with a circular top-hat filter of radius one pixel; after smoothing, we classified as `halo' particles all corresponding pixels with a value $>0.99$.}

Initially, the angular momentum has a clear quadrupolar pattern in the centre (Panel A of Fig.~\ref{fig:2d_sim}), summing to zero around circles. At the outskirts, though, the boundary of the protofilament is not circular, giving a net angular momentum (Panel B) \citep{Doroshkevich:1970}. As this matter falls onto the filament, pulled by gravity, it carries this angular momentum to the filament axis. Collapsed particles appear randomly arranged, losing their initially obvious quadrupole (Panel C). Here, `collapsed' particles may not be virialized; they simply experience multistreaming along two orthogonal axes. In Panel D, we quantify the evolution of the monopole for a stacked sample of 183 such haloes in the simulation. As time passes, the averaged angular momentum is transported from the outskirts toward the centre. Collapsed particles contribute most to the angular momentum near the centre of the filament, whereas non-collapsed particles dominate the angular momentum at the outskirts. Near the filament centre, the ratio between the monopole and quadrupole is zero initially, but then increases with time as the monopole increases and the quadrupole diminishes. In summary, angular momentum around filaments originates according to the tidal-torque theory \citep{peebles:1969,white:1984}: an asymmetric matter distribution is torqued up and is later transported gravitationally to the centre. As with dark-matter haloes \citep{white:1984,motloch/etal:2020}, the angular momentum of a filament cross-section seems to grow as the scale factor $a^{3/2}$ until collapse, after which angular momentum is conserved in physical coordinates, generally retaining its direction \citep{neyrinck/etal:2020}.

\section{Analysis in terms of physical filament properties}
\label{sec:indepth}
\mnedit{Here, we investigate filament spin in greater depth, particularly important because of our empirical, rather than physical, filament definition. First, in
\S\ref{sec:visualquality}, we look at how these filaments defined simply as regions between rather large haloes correspond to what we expect visually.
\S\ref{sec:densitysandwich}, we look at how the spin correlates with the density and arrangement of matter in cylindrical filters such as used in our filament definition.

In the rest of the section, we use alternative simulations, practically because they are accessible for analysis with the \origami\ web-type classifier that we use, but also this broadens the range of cosmological settings of our results. In \S\ref{sec:origamiwdm}, we look at the degree of physical collapse in empirical filaments. \S\ref{sec:filamentcoherence} examines the amount and coherence of spin along collapsed density ridges between haloes, rather than along straight lines. For this analysis, we use both the \origami\ and MMF2 filament-finders to find these `collapsed density ridges.'

Examining rotating filaments in terms of the \origami\ and MMF2 definitions, in addition to our fiducial definition, should sufficiently elucidate the spin's physical nature and definition-dependence for the purposes and scope of this paper. But there are several other well-motivated filament finders in common use that are worth mentioning. A non-exhaustive discussion of them follows; for a more comprehensive comparison, see \citet{libeskind/etal:2018}.

Several filament-finders work with the density field or the point process itself of galaxies or haloes; this is the most straightforward class of filament-finders to apply to observed galaxies. {\scshape DisPerSe} \citep{sousbie:2011} has already been used to study the vorticity and spin field around filaments \citep[e.g.][]{PichonEtal2011,codis/pichon/pogosyan:2015}. It finds density ridges, conceptually-similarly to MMF2 \citep[see also the original MMF,][]{aragon-calvo/etal:2007} and {\scshape nexus} \citep{cautun/etal:2013}. But {\scshape DisPerSe} works on the  scale of individual particles, rather than on the field as smoothed on multiple scales, which can cause results to differ somewhat. The Bisous filament-finder \citep{tempel/libeskind:2013} also works with galaxies as points, finding strings of points commonly appearing together in cylindrical filters; Bisous filaments are broadly similar to density ridges.

Another, dynamical class of algorithms brings in further information beyond the density. \origami\ is this category, using ordering differences between particles’ Lagrangian (initial) coordinates and Eulerian (late-time) coordinates. Others use the tidal force field configuration \citep{hahn/etal:2007,Forero-RomeroEtal2009}, or the velocity shear field \citep{HoffmanEtal2012}. See \citet{ganeshaiahveena/etal:2018} for some discussion of how filaments defined by the density and velocity shear field relate to each other.}

\subsection{Visual filamentarity between Millennium halo pairs}
\label{sec:visualquality}
\mnedit{Although the density stack in Fig.\ \ref{fig:isotropic} of 33,951 empirical filaments (defined as regions between two large haloes) looks quite like an idealized filament, we wondered how they individually correspond to visual expectation. \citet{colberg/krughoff/connolly:2005} found by visually inspecting halo pairs in this length range that they are joined by a recognizable `straight or warped' filament $86\%$ of the time; however, they used haloes with mass threshold $10^{14}\hsolar$; with our smaller $10^{13}\hsolar$ cut, we expect the fraction of recognizable filaments in our sample to be somewhat smaller.

Here we show flyarounds of the density fields around four example halo pairs from our sample, spanning a range of filamentarity:
\href{ttps://qx211.github.io/assets/movie/Filament_15366.mp4}{1};
\href{https://qx211.github.io/assets/movie/Filament_31267.mp4}{2};
\href{https://qx211.github.io/assets/movie/Filament_2355.mp4}{3};
\href{https://qx211.github.io/assets/movie/Filament_27901.mp4}{4}. Unsurprisingly, looking at several examples, we saw many cases where the pair of haloes are connected by a matter bridge close to a straight line, but also many cases otherwise. Nonetheless, it is clear that on average, there is a matter bridge in our sample (Fig.\ \ref{fig:isotropic}), and they rotate substantially.} 

\subsection{Dependence of spin on matter arrangement in a cylinder}
\label{sec:densitysandwich}
\begin{figure}
    \centering
    \includegraphics[width=1.01\columnwidth]{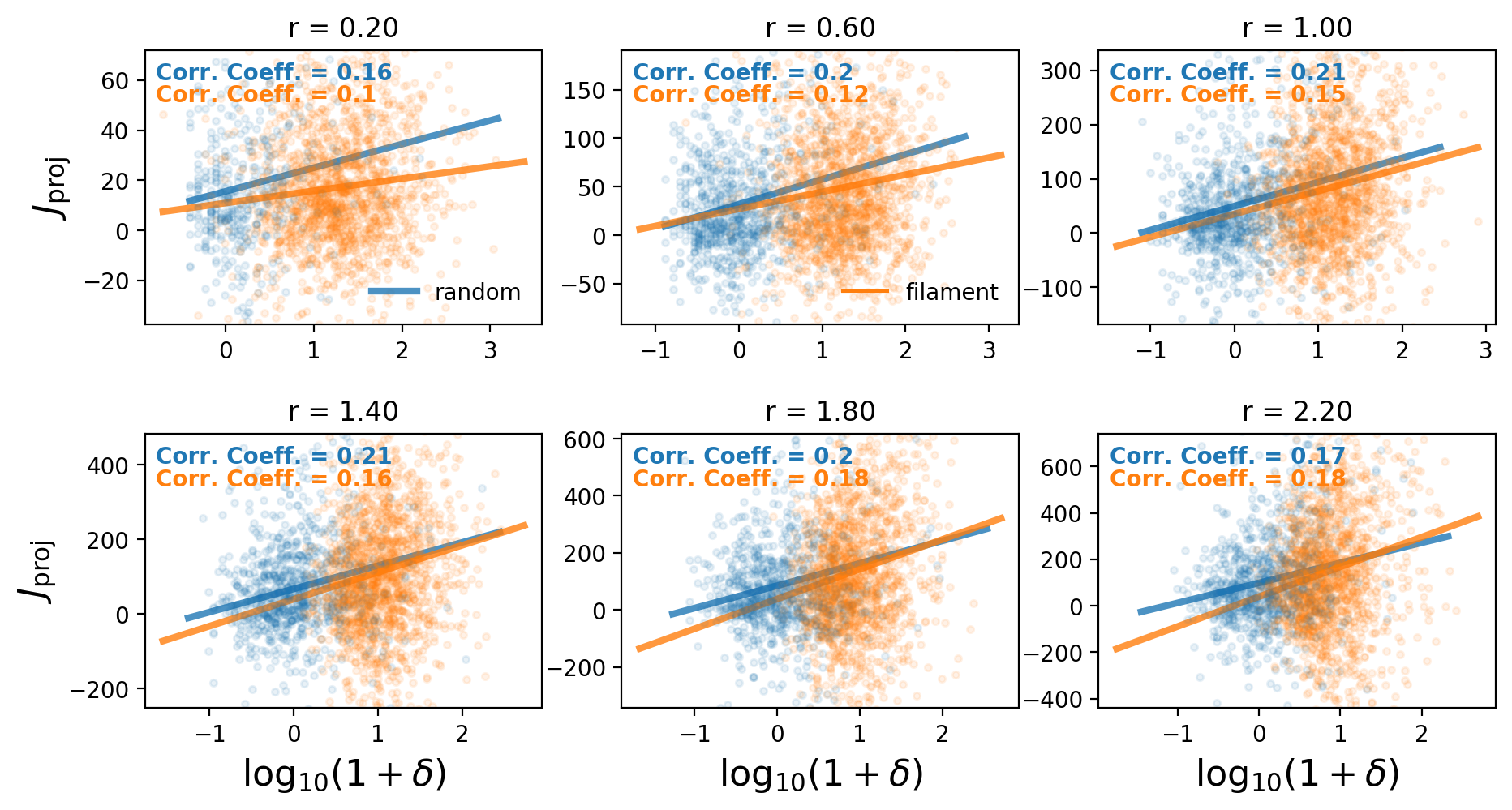}
    \caption{The dependence of the angular momentum per particle on the density within cylindrical `filament region' filters, for six radial bins (indicated by titles above each plot). The density-rotation relation for random cylinders (one point per cylinder) is shown in blue, and for our sample of halo-bracketed filaments in orange. Dashed lines show best-fitting linear models.}
    \label{fig:mass_dependent}
\end{figure}

\begin{figure}
	\centering
	\includegraphics[width=0.666667\columnwidth]{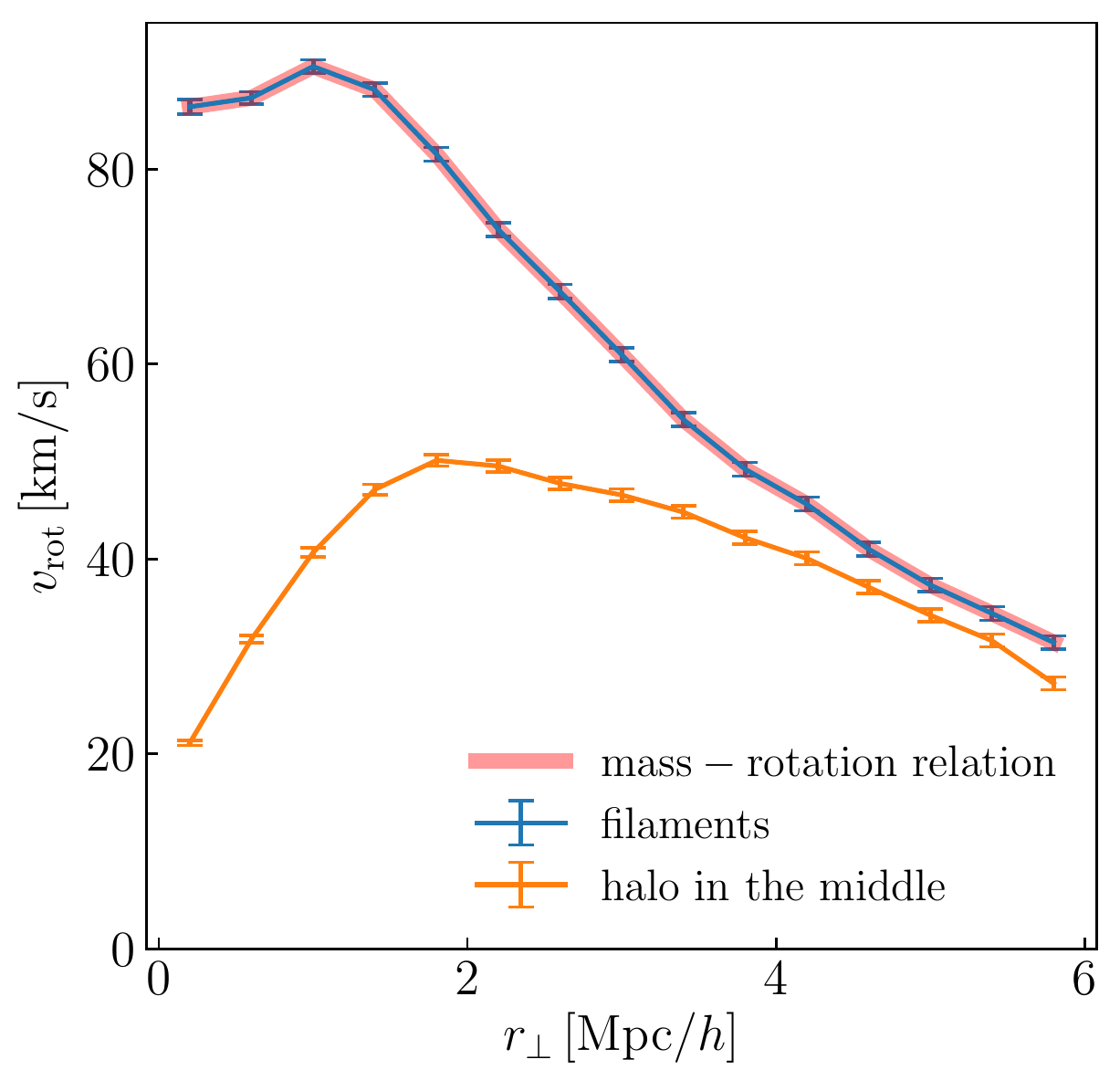}
	\caption{The rotational velocity per particle as a function of distance to the filament axis. The blue line shows the result from pairs of haloes that are separated by 6-10 $\hmpc$. The orange line shows the result from random pairs of points with this range of separation, but with a halo of at least $10^{13} \hsolar$ in the middle.The pink line shows the expected value based on the mass-rotation relations found in Fig.\ \ref{fig:mass_dependent}.}
	\label{fig:sandwich}
\end{figure}

As one way to elucidate the physical nature of the filaments in our fiducial MS sample, we look at a filament's rotation speed depends on its average density. We do so by splitting our filament sample according to the mass density averaged within the cylindrical filament region, $\delta$. As shown in Fig.~\ref{fig:mass_dependent}, the rotational velocity $v_{\rm rot}$ increases with $\delta$. This is expected from the gravitational origin of the rotational velocity, since filaments with more mass have brought matter from larger distances, with a longer lever arm in their angular momentum sum. In Fig.~\ref{fig:sandwich}, we show that by using the best-fitting relation between $v_{\rm rot}$ versus $\delta$, the rotational velocity is accurately recovered.

Curiously, though, Fig.~\ref{fig:mass_dependent} shows a density dependence even for randomly selected line segments. Is this all we are seeing in our results, that our filaments are just particularly dense regions using our cylindrical filter? We investigate the effect of the shape of the mass arrangement in Fig.~\ref{fig:sandwich}. We select random `filaments' between pairs of points centred on random haloes. Typically, this matter distribution would be far from filamentary. In the figure we find that the rotational velocity of the non-filamentary matter distribution is much weaker than the case of filaments. This suggests that it is important to have an actual filamentary matter distribution to give substantial rotation.

\begin{figure}
    \centering
    \includegraphics[width=0.66667\columnwidth]{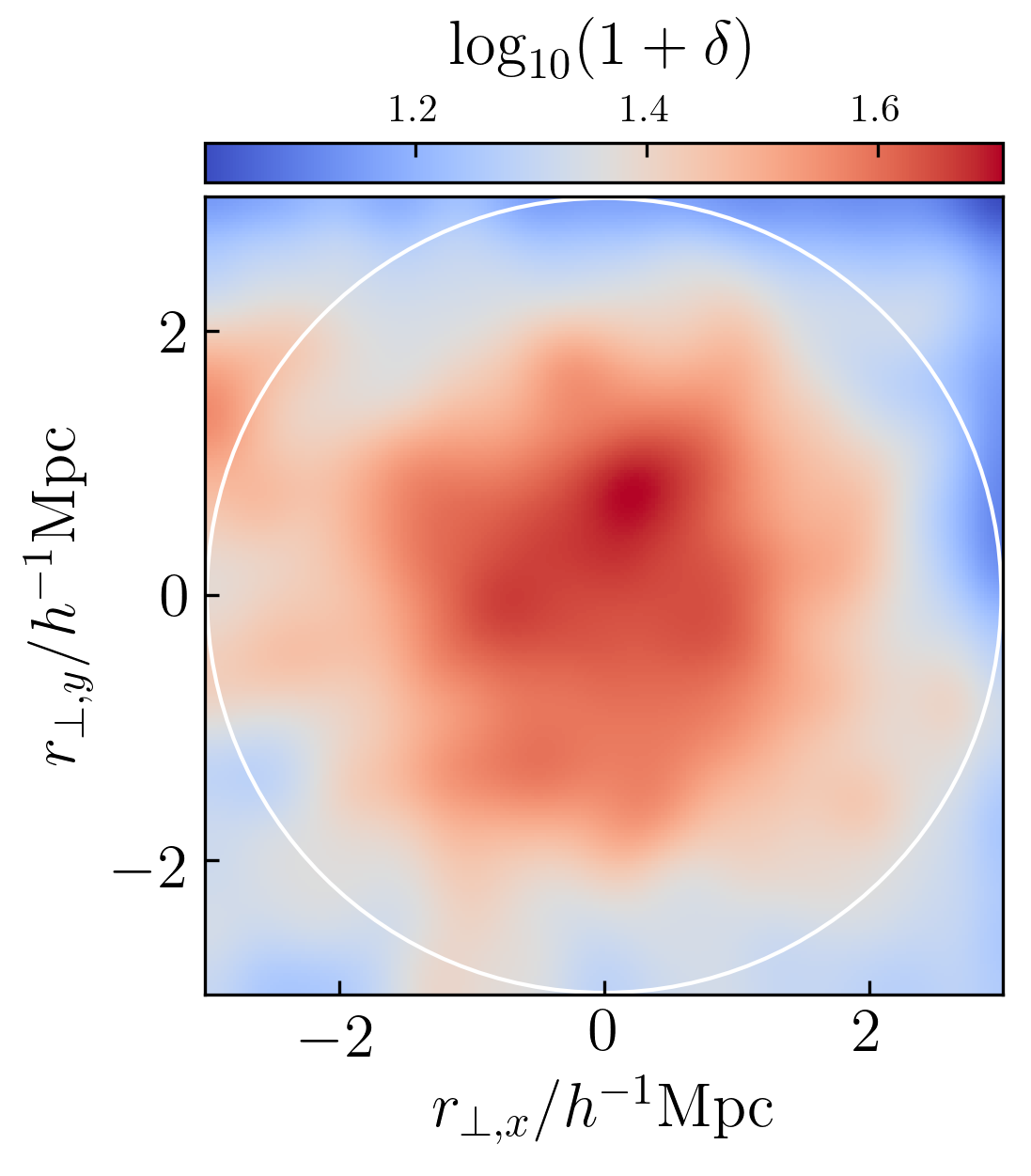}
    \caption{The average matter-density cross section of the top 10\% densest random line segments, as measured in the inner 1 $\hmpc$.}
    \label{fig:random_cross_sec}
\end{figure}

In another test, we look at a projected cross-section of random cylinders in the top 10\% of the density range. In Fig.~\ref{fig:random_cross_sec}, both the selected filaments and random line segments are on average cylinder-like structures, as in Fig.~\ref{fig:isotropic}A. Whether we select cylinders that lie between a pair of haloes, or simply mass within that cylindrical filter, we tend to pick up filamentary regions. In this sense, our definition of filaments is rather general, and we expect the results to hold for a wide range of filament definitions that detect elongated mass distributions.

\subsection{Relation to a collapsed-filament definition}
\label{sec:origamiwdm}
\mnedit{Here we study an aspect of our filament definition, exploring the contribution of the spin from high-density regions that have undergone cylindrical collapse. In an idealized picture, the spin would come from such physically collapsed filaments, but with our empirical definition, we expect that instead, some halo pairs do not have collapsed density ridges between them.

There are various possible degrees of `collapse.' An idealized filament would have experienced cylindrical collapse, along two directions perpendicular to its axis (its particles classified as `filament' by \origami, which we use for this test). Realistically, most would contain haloes that have collapsed along all three axes as well, classified as `halo' by \origami. We count these as part of the filament. Such fragmentation would be entirely expected for even an idealized filament in the MS, since small-scale fluctuations typically drive small haloes to form in such a filament. We also wish to test to what degree a filament sample like ours contains wall or even void regions.}

We could not apply this analysis to the MS because \origami\ cannot analyse a simulation with glass initial conditions (like the MS). Instead, we use a warm-dark-matter (WDM) dark-matter simulation from \citet{yang/etal:2015}, which assumed $\Lambda$CDM cosmological parameters slightly different from the MS. Smoothing the initial conditions produces a more visually prominent cosmic web \citep[see Fig.\ 3 of] []{neyrinck:2012}. The 100\,$\hmpc$, 512$^3$-particle simulation we use had initial conditions smoothed at $\alpha=0.1$\,$\hmpc$ (but with a gradual kernel, extending a factor of $\sim 6$ larger in scale).

\begin{figure}
    \centering
    \includegraphics[width=\columnwidth]{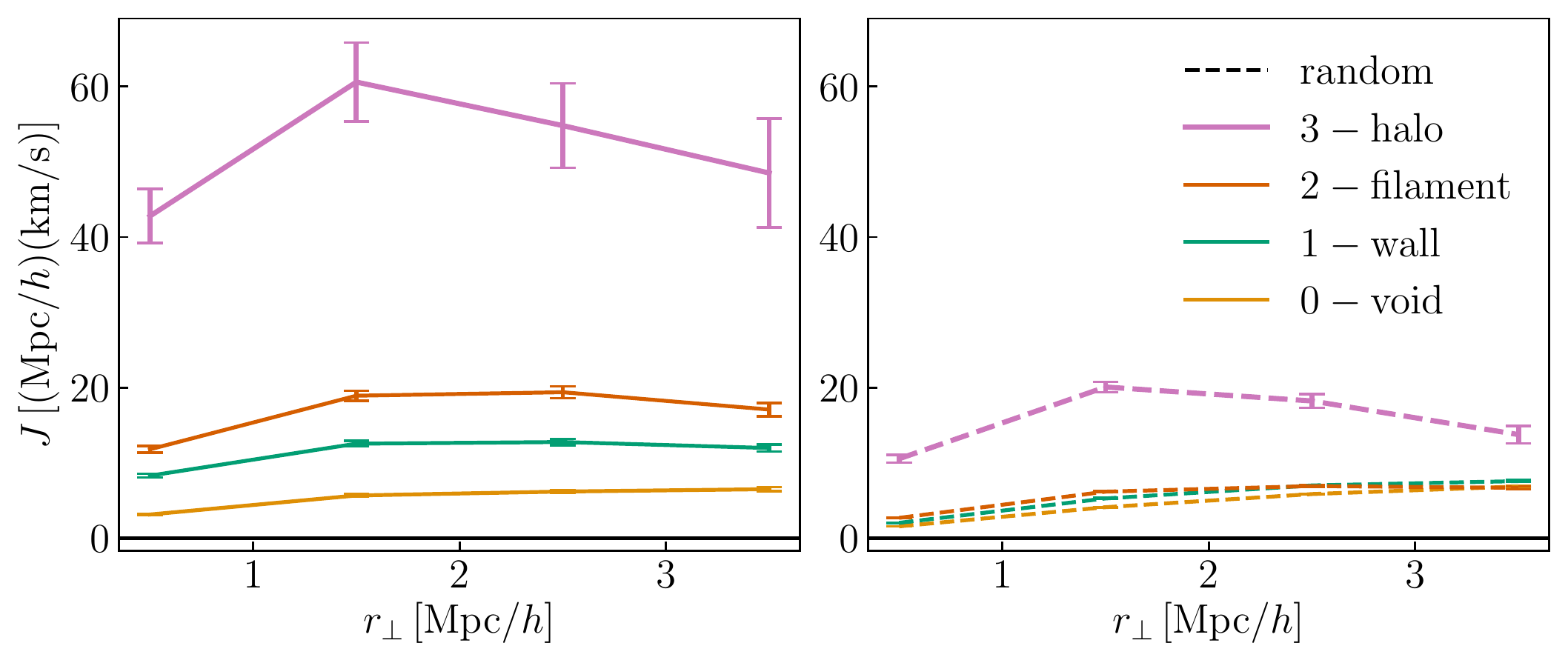}
    \caption{Angular momentum density profile $J(r_{\perp})$ contributed from particles with different origami tags in the WDM simulation, for filaments 6-10\,$\hmpc$ long (left) and for a random sample with the same length (right). The tags indicate the number of axes of collapse; i.e., 0 (void), 1 (wall), 2 (filament -- according to \origami, not our straight-line definition), and 3 (halo).  $J(r_{\perp})$ increases as the degree of collapse increases.
    Error bars indicate the error on the mean.}
    \label{fig:wdm_tags}
\end{figure}

We identified the haloes used to define filaments in the 3D WDM simulation in a new way, similar to that used for the 2D simulation: we first tagged particles with \origami\ \citep{falck/neyrinck/szalay:2012}. We joined together `halo' particles that were adjacent on the Lagrangian grid. We first applied an erosion operator with a radius of one grid spacing, to cut some visually spurious links, and to remove tiny spurious haloes within filaments \citep{wang/white:2007}. We used this method because the outer (`splashback') caustics of haloes in WDM are more visually evident than in CDM, and the \origami\ method worked well to identify these halo boundaries \citep{neyrinck:2012}. This fully Lagrangian method may be promising more broadly as a halo-finder, but is especially useful for a WDM simulation with a clarified cosmic web. The haloes we used had at least 1000 particles, resulting in a sample of 1825 haloes. Because of the substantial differences in the halo finder, mass estimate, and WDM cosmology, we do not treat this halo sample as equivalent to our MS sample, but these results are still relevant.

In Fig.\ \ref{fig:wdm_tags}, we see that for filaments in the WDM simulation (left), the signal is dominated by halo particles, with decreasing contributions from filament and wall, and totally uncollapsed void particles. Note that the `halo' particles are not in the haloes at the end of the filament, but are in (likely smaller) haloes between them. Also, because of spurious fragmentation in WDM filaments \citep{wang/white:2007}, some filament particles may have been classified as halo particles. So, we consider both `halo' and `filament' particles to be filament matter that has collapsed. Another reason to consider \origami\ `halo' particles to be collapsed along the filament is for near-invariance to a small-scale power cut-off. If this were a CDM rather than WDM simulation (e.g.\ in the MS), the filament would have fragmented into haloes. At right, we do the same measurement, except using random filament endpoints instead of haloes. In that case, the contribution from each type is much smaller.

\subsection{Sensitivity to filament definition, and lengthwise rotation coherence}
\label{sec:filamentcoherence}
Here, we take this analysis further, tracking filament spin closely along collapsed density ridges between haloes, almost never straight.

\citet{PichonEtal2011} made a related measurement, in a paper focusing on how filaments contribute to halo and galaxy angular momentum. In their Appendix A, they show a correlation function of angular momentum vectors along {\scshape DisPerSe} filaments, finding that angular momentum along filaments is highly correlated on scales up to 15 $\hmpc$.

\subsubsection{Method}
For this measurement, we analysed filaments identified in a dark-matter-only re-run of the Illustris \citep{nelson/etal:2015} simulation, assuming cosmological parameters from the Nine-Year Wilkinson Microwave Anisotropy Probe analysis \citep{HinshawEtal2013}. We did not expect qualitative differences in our results from this slight change in cosmological parameters, but it is good to check. In our re-run of Illustris, we smoothed its initial conditions with a 2\,$\hmpc$ sharp-$k$ filter. We could have used this Illustris-rerun above in \S\ref{sec:origamiwdm}, but we used both, to explore results in a variety of simulations and halo-finding methods (there, \origami, applied directly to a WDM simulation, and here, using large haloes identified with FOF in the unsmoothed Illustris simulation).

To use \origami\ in this context, it was necessary to infer a set of grid (instead of glass) initial conditions from Illustris-3-Dark. We formed a regular $512^3$ grid by interpolating the velocity fields from the first publicly available snapshot ($z=46.77$), using a Delaunay-based interpolation scheme \citep{vandeWeygaert/schaap:2009}. We then recovered grid initial conditions using the inverse Zel'dovich approximation to recover the Illustris initial conditions. We checked that by eye, the large-scale cosmic-web structure in our remapped runs and the original Illustris-3-Dark correspond.

We generated an MIP ensemble \citep{aragon-calvo:2016} of 32 realizations from the smoothed initial conditions, allowing negligible discreteness noise. The initial conditions of the 32 realizations share the same Fourier modes above a 2\,$\hmpc$ cutoff scale, but are populated with different modes on smaller scales. Using particles from all realizations of this correlated ensemble allowed us to compute velocity fields with negligible discreteness noise for each filament, even with a naive cloud-in-cell velocity estimation method.

We applied two different web classifiers to identify the collapsed density ridge: MMF2 (Multiscale Morphology Filter 2, based on the density Hessian, applied at multiple scales) and \origami\ (a dynamical definition using directions of particle crossings).
There were 86 pairs of haloes of mass at least $10^{13} \hsolar$ that satisfied our separation cut of 6-10 $\hmpc$. We found that 32 of these were separated by clear bridges of MMF2 filament-classified voxels; all of these had clear bridges of \origami\ particles. There were also typically small haloes (and particles \origami-classified as `halo') between the endpoints; we consider any haloes between the endpoints to be part of the filament, and include their particles in the analysis.

We grouped particles into 7 quantiles along the `filament region' (Fig.~\ref{fig:illustration}) of each filament, measuring the spin within each segment. We positioned the quantiles to give (almost) the same number of particles in each segment, rather than an equal-length division.

\begin{figure}\centering
    \includegraphics[width=\columnwidth]{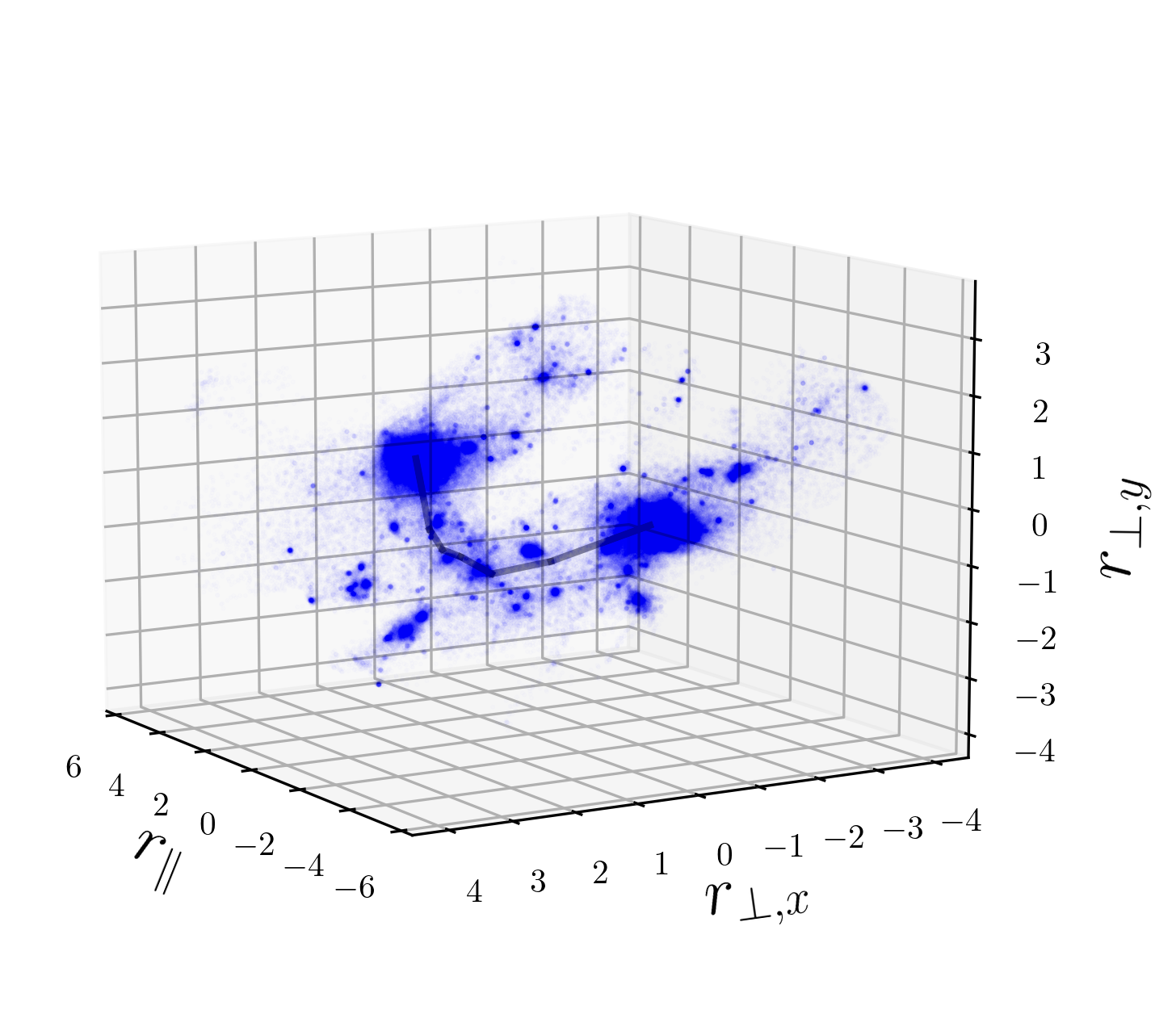}
    \caption{Segmentation of a filament in our re-run of the Illustris simulation. Particles tagged by \origami\ as filament and halo particles are coloured blue. The thick black curve shows the 6 segments of this filament. A fly-around animation of this example is available at \url{https://qx211.github.io/img/Filament_65.mp4}.}
    \label{fig:demo}
\end{figure}

We traced the ridge by linearly interpolating between nodes, one node per quantile. We positioned each node in $x$ and $y$ (if the $z$-axis is along the straight line separating endpoints) by finding the median $x$ and $y$ particle coordinates in that quantile (neglecting particles that have distance larger than 3 times the standard deviation in that bin). A `segment' occupies the space between adjacent nodes; there are 6 segments between the 7 nodes of a filament. In each segment, we measured the angular momentum around an axis connecting the nodes that bracket it. An example of the segmentation is shown in Fig.~\ref{fig:demo}.

We then measured 2D rotational velocity profiles for each filament segment, setting the positive direction to align with the filament's rotation within the inner 1 $\hmpc$, averaged among all segments. To achieve negligible particle discreteness noise, we stacked particles over 32 MIP realizations. For each of the 2D rotational velocity maps, $v_\theta(r,\theta)$, we measure the amplitude of the quadrupolar pattern at different phases $\phi$ at distance bin $r_i$ via
\begin{align}
    Q(r_i,\phi) = \int_0^{2\pi}\int_{\rm r\in r_i} v_\theta(r,\theta)\cos(2\theta + \phi) r dr d\theta,
\end{align}
and we define the phase of the quadrupolar pattern to be the value $\phi_1$ of $\phi$ that maximizes the quadrupole ampliude $Q$, measured from pixels within 1 $\hmpc$ of the filament axis. At each radius, we set $J_2(r_i)$ at radius bin $r_i$ to be the amplitude of the quadrupole with phase $\phi_1$ measured in the rotational velocity field $v_\theta(r_i,\theta)$, including pixels within the radius bin $r_i$. Note that even though $J_2(r<1\hmpc)$, averaged over the inner $1 \hmpc$, is positive, $J_2(r)$ can go negative at larger radius, if the phase of the quadrupole at that radius differs sufficiently.

\subsubsection{Results}
\begin{figure*}
    \centering
    \includegraphics[width=\textwidth]{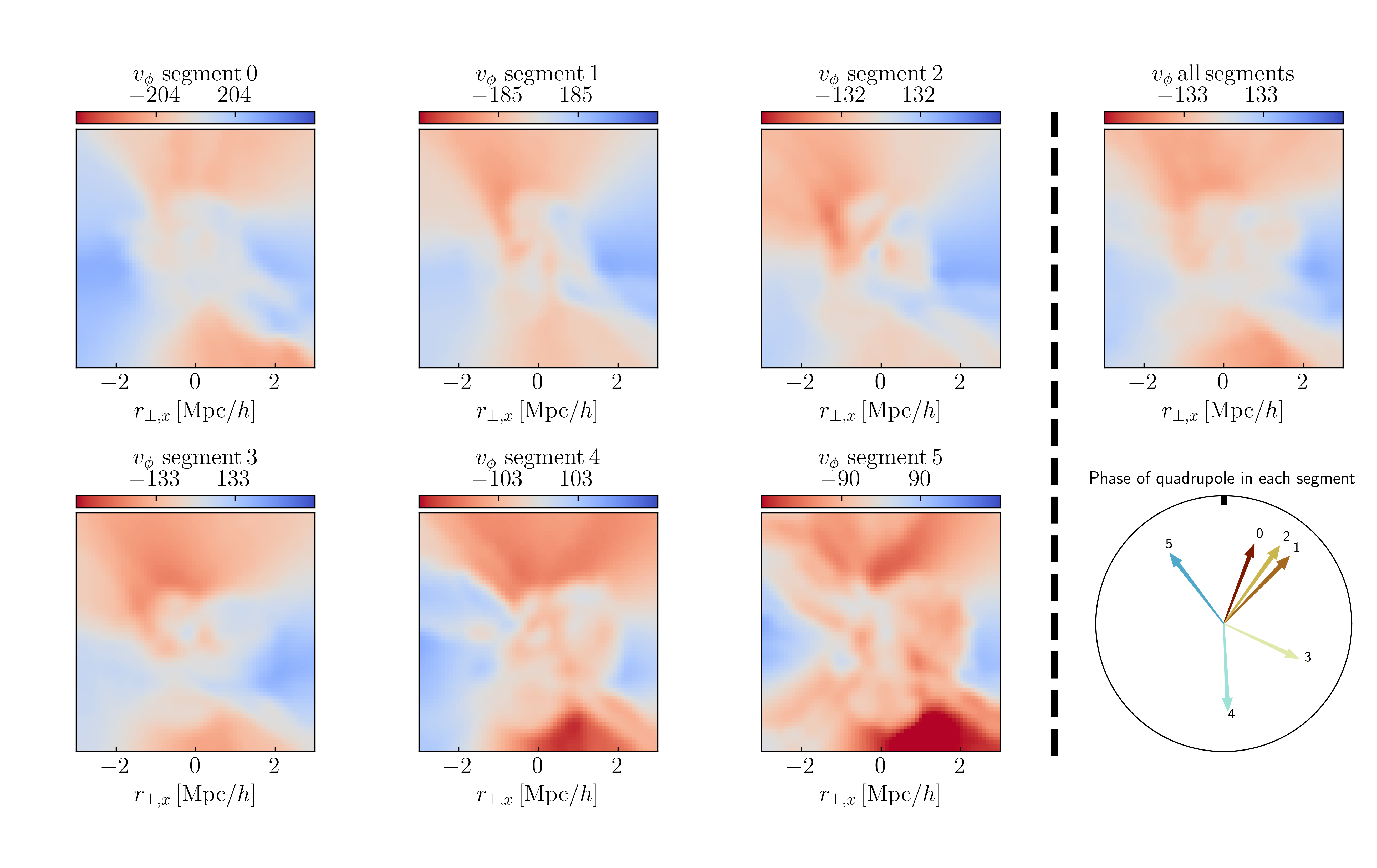}
    \caption{Left: Average rotational velocity field, in km/s for each of the six segments of a filament. Right: The average rotational velocity profile from combining particles in the six segments together. The clock plot in the lower-right panel shows how the quadruople's phase changes across segments. 12 o'clock is defined as the phase estimated from the entire filament.}
    \label{fig:2D_vol}
\end{figure*}

\begin{figure*}
    \centering
    \includegraphics[width=\textwidth]{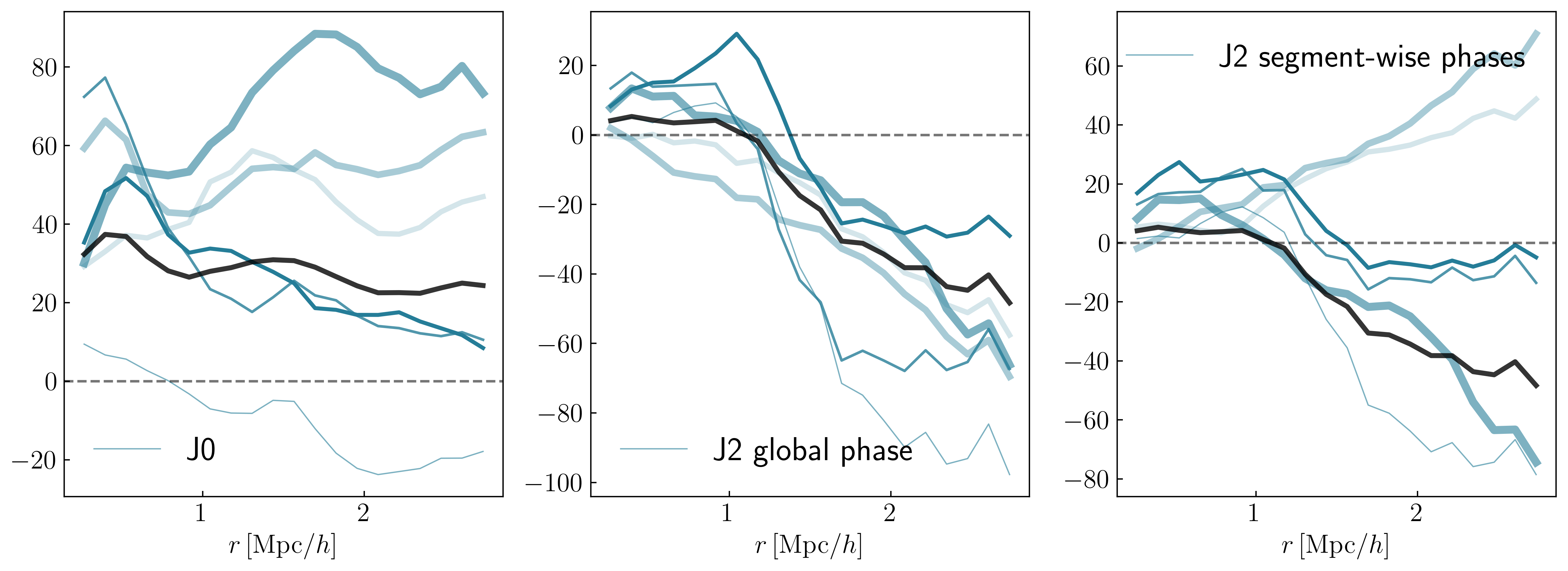}
    \caption{Monopole $J_0(r)$ and quadrupole $J_2(r)$ from the velocity (in km/s) maps shown in Fig.\ \ref{fig:2D_vol}. Blue line thicknesses and opacities increase with segment number. The phase giving the maximum amplitude for $J_2$ is set in the innermost 1 $\hmpc$, and then is used to measure $J_2$ at all radii. It goes negative where the quadrupole's phase is sufficiently rotated from the inner region. In the middle panel, this phase is set by the $v_\phi$ averaged over all segments. In the right panel, the phase is determined segment-wise, i.e.\ to maximize $J_2$ in the inner 1 $\hmpc$ of each segment. This is why the right panel generally has higher amplitude than the middle.}
    \label{fig:vel_multipole}
\end{figure*}

An example filament is shown in Fig.~\ref{fig:2D_vol}; a 2D rotational velocity map is shown for six segments. The quadrupolar (and more general anisotropic) pattern is indeed visually evident in this case. From the clock plot, we can also see the rotation of the quadrupole among the six segments, i.e. the direction of the quadrupole rotates from segment to segment along the filament. In Fig.~\ref{fig:vel_multipole}, we can see that the amplitudes of the monopole and quadrupole in each segment vary, i.e. $J_0$ can be larger or smaller than $J_2$. But even though visually, what stands out is the anisotropic pattern, the monopole is dominant at low radius, in most segments. Once averaged over the entire filament, $J_0$ is generally dominant over $J_2$, which we show in Fig.\ \ref{fig:pca}D, for easy comparison with the Millennium results. This monopole can be compared to the MS filament velocity profile in Fig.\ \ref{fig:isotropic}D, which is larger; this is counter-intuitive if we expect the collapsed ridge (which we trace in Illustris) to have a greater speed. The small sample size (only 32 filaments) makes this a weak statement, however.

\begin{figure}
    \centering
    \includegraphics[width=\columnwidth]{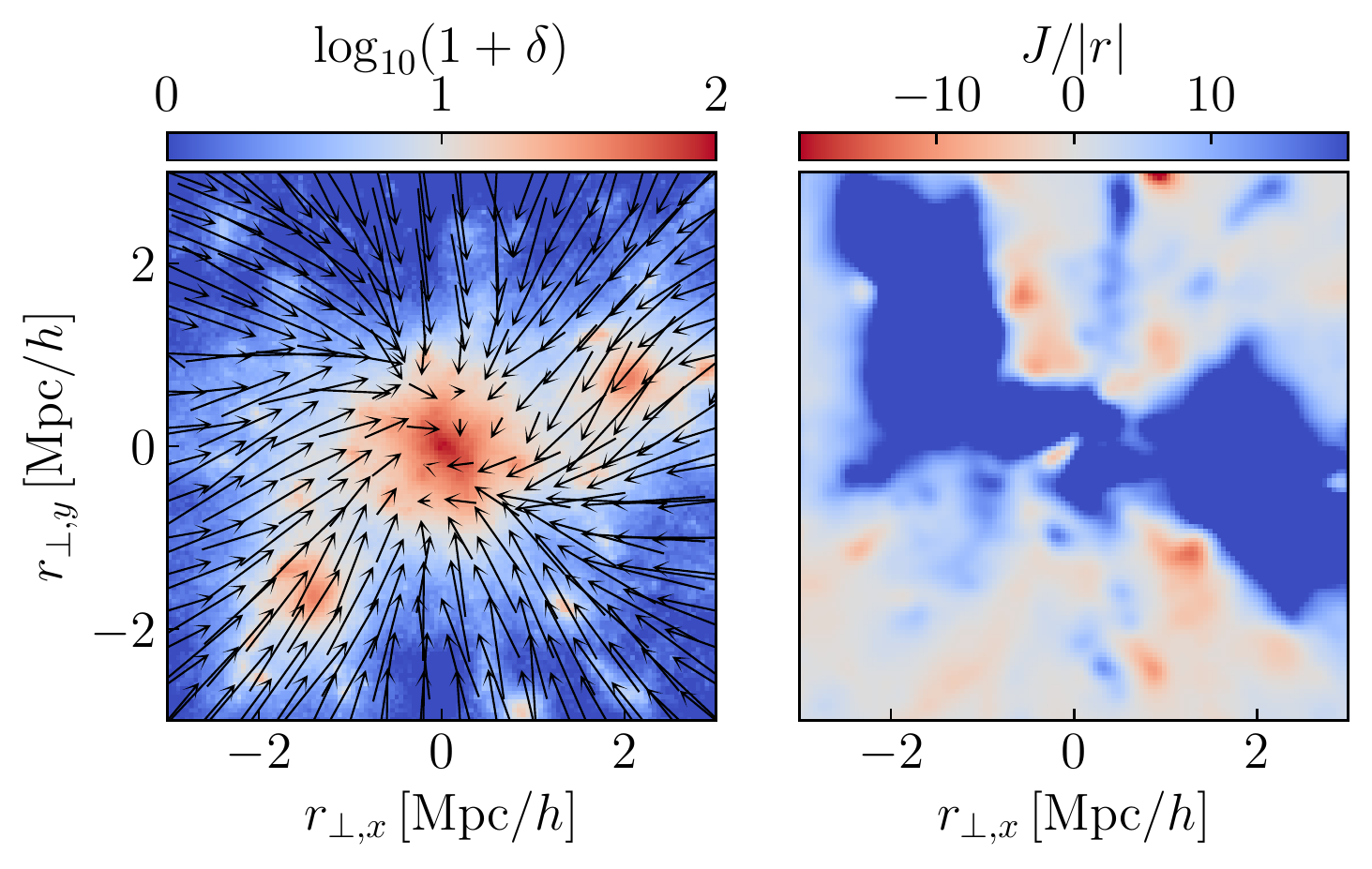}
    \caption{Similar to Fig.~\ref{fig:pca} A and C, but for a much smaller sample of 32 Illustris filaments. Left: the stacked density, with arrows representing velocities. The maximum length of a vector corresponds to 1176 km/s. Vector lengths $\ell$ were transformed using $\ell= 50\,\sinh^{-1}(v/50)$, with $v$ in km/s; e.g.\ a vector representing 50 km/s is 4.4 times shorter than the maximum. Right: the azimuthal component of the angular momentum. In both stacks, each of the 6 segments along each filament has been rotated so that the PCA major axis of the density distribution aligns with the horizontal axis. Deep blue patches at right correspond to large clockwise velocity arrows at left.}
    \label{fig:2DTangential}
\end{figure}

In Fig.\ \ref{fig:2DTangential}, we show average flows onto filament collapsed density ridges (at center) over this 32-filament sample, rotating the signal in each of the 6 segments to align as in Fig.\ \ref{fig:pca}. The results are qualitatively the same as in that figure, but the horizontal `wall' in the density field is hardly evident, because of the much smaller sample. The velocity field displays similarity to Fig.\ \ref{fig:pca}C, though.

\begin{figure}
    \centering
    \includegraphics[width=\columnwidth]{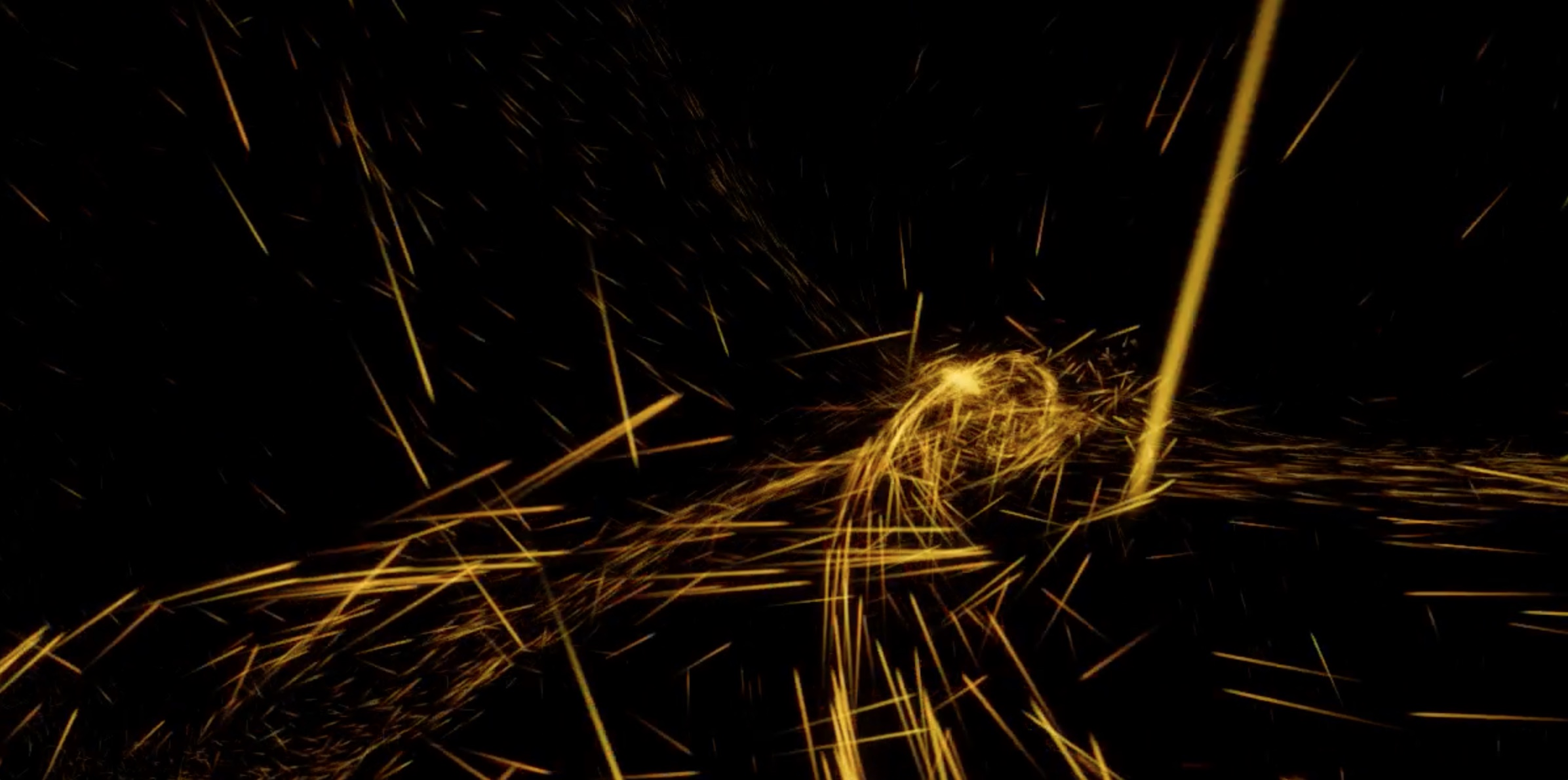}
    \caption{A view of particles (golden comets) advected along streamlines of the final snapshot of the velocity field in our smoothed re-run of the Illustris simulation (starting the animation after some advection has already happened). We highly recommend watching our animation showing another view of the particles being advected, at \url{https://www.youtube.com/watch?v=h1-a-htHAxY}}
    \label{fig:rotatingfilament-frame}
\end{figure}

We show the velocity field of a $\sim 4 \hmpc$-long filament from the Illustris simulation in Fig.\ \ref{fig:rotatingfilament-frame}. An animation of advected particles in another view is linked in that caption. What is only a hint of a helical loop in this frame is a much clearer helical flow in the animation. The velocity field in this animation is an average over DTFE \citep{vandeWeygaert/schaap:2009} velocity fields from 32 Illustris MIP realizations, corresponding to an initial power truncation at 2 $\hmpc$ (explained above). The velocity field was further high-pass-filtered with a 0.5 $\hmpc$ Gaussian, to remove large-scale flows. We used the Unity engine for the advection and rendering.

\subsubsection{Summary}
In this subsection, we used another standard reference simulation (Illustris, with many different realizations of small-scale modes), and two different filament definitions, going beyond our fiducial straight-line definition. By studying individual filaments in depth, we saw that the amplitude of the monopole and quadrupole along the filament are generally comparable, with the phase of the quadrupole shifting. When averaging over the entire filament, because of this phase rotation, the averaged quadrupole is typically suppressed compared to its value in single segments. So a straight-line filament definition may overestimate the dominance of the monopole over quadrupole, but qualitatively, the results are the same: the spin-field monopole is generally of comparable or higher magnitude than the quadrupole.

\section{Candidates for the longest rotating objects in the Universe}
\label{sec:object}
Here we clarify and elaborate on our statement that the longest substantially rotating objects in the Universe are likely filaments. We word this carefully; we do not mean that all filaments are substantially rotating, and are longer than all other substantially rotating objects in the Universe. Instead, we mean that the few longest rotating objects are likely to be filaments.

An `object' here is a cosmic-web component, a coherent virialized or collapsed object (halo, filament, or wall), or a region without collapse (void). For definiteness, we assume a dynamical dark-matter-sheet folding definition of these objects, as in \origami\ \citep{falck/neyrinck/szalay:2012}, but some other definitions broadly agree, e.g.\ the MMF2 \citep{aragoncalvo/yang:2014}, and {\scshape nexus} \citep{cautun/etal:2013}; see \citet{libeskind/etal:2018}. In the \origami\ stream-crossing definition, a supercluster, an overdensity in the density field smoothed on at least tens of $\hmpc$, would likely contain many virialized haloes, voids, collapsed filaments, and collapsed walls.

A filament typically originates from an oblate, disk-like object in the initial conditions \citep{zeldovich:1970,ShandarinZeldovich1989,falck/neyrinck/szalay:2012,LovellEtal2014,HiddingEtal2016,FeldbruggeEtal2018}. The oblate object turns prolate, stretching along the axis that becomes the filament axis, and compressing toward it. As with a halo, any small angular momentum in the collapsing protofilament leads to faster rotation, like a figure skater pulling in their arms. 

But there is no equivalent process in the formation of a void or wall that is likely to speed up rotational motions. Walls do collapse along one axis, and can contain rotating patches (haloes and filaments within them). However, they are unlikely to rotate coherently (e.g.\ as a disk), since they generally expand faster than the cosmic mean in the two directions perpendicular to the collapse axis \citep{aragon-calvo/etal:2011}, evolving similarly to voids in a 2D Universe. This serves to suppress rather than enhance rotational motions.

Voids are even less likely to rotate coherently. They are `cosmic magnifiers' \citep{aragon-calvo/szalay:2012}, their interiors expanding faster than the cosmic mean, resulting in vanishing vorticity \citep[e.g.][]{hahn/etal:2015}, and negligible large-scale rotation expected.

As for haloes, they are well-known to spin, but are of maximum radius (e.g.\ as measured from the Millennium haloes) $\sim2$\,$\hmpc$, and so do not compete in size with filaments.

Another, mere suggestion for filaments as the longest objects is that in the toy, origami approximation \citep{neyrinck_iau:2016,neyrinck/2016}, filaments, which can be arbitrarily long, are the largest rotating components; walls and voids do not rotate. Voids do not rotate by construction in this model, but the non-rotation of walls is not trivial.

The situation in the real Universe is more ambiguous, of course. While spin along filaments has substantial coherence, this coherence degrades with distance. So the longest filaments observed may have only slightly aligned spin at both endpoints, so the most extreme objects would likely be open to interpretation.

Also, patches exist that straddle the clear categories of wall, filament, and halo. Boundaries between objects (or objects and sub-objects) can be ambiguous, as well. While we have argued that walls and voids are unlikely to rotate substantially, it would be interesting and relevant to measure the degree of wall and void rotation in simulations, although their complex geometries would make this a more subtle measurement than for filaments.

\section{Discussion}
\label{sec:conclusion}
We find that cosmic filaments generally carry some spin, which has some coherence along them. Within the filament, the dark matter has substantial random and radial motion, though; one should not think of filaments as nearly uniformly rotating cylinders. The velocity field in a filament consists of overlapping streams of dark matter going in different directions, with a net rotation after averaging over all streams. One might hesitate to call this rotation, but in fact this is the same situation as with haloes, which are said to rotate, despite even more severe multistreaming. We also should clarify that there is typically much substructure along each filament; no individual filament looks as idealized as the stack in Fig.\ \ref{fig:isotropic}.

As with halo spin, we advocate thinking of filament spin as a fundamental property, even though that spin can be small compared to random motions, and a practical spin measurement for a filament (as with a halo) depends on details and definitions. But we must also acknowledge that even though it is generally a dynamically younger object, the velocity field of a filament can be more complex than of a halo. A halo is readily approximated by a single velocity and a spin, while in principle the spin along a filament can vary with position along it (typically mildly, as we have shown), and the filament's shape could distort, something that we have not measured here.

\mnedit{During the review process of this paper, we learned that another team \citep{wang/etal:2021} recently claimed detection of a sample of rotating filaments outlined by galaxies in the Sloan Digital Sky Survey, prompted by our present work. Still, we give our own thoughts about detecting filament rotation in the observed Universe here. The most straightforward way to detect rotation in filaments} may be to identify thick filaments of galaxies between clusters; one could look for rotation in the galaxies and gas between them such as using radio surveys to locate turbulent gas in intra-cluster bridges \citep{brunetti/vazza:2020}. For a rotating filament nearly in the plane of the sky, on average, one side will move away from, and the other towards, the observer, as in Fig.\ \ref{fig:isotropic}B, giving a redshift and blueshift, respectively. That is, redshift-space distortions would tend to shear, or tilt, the filament along the axis, such that the side moving toward the observer is closer. This might be detected by stacking galaxy redshift-space distances along filament axes, and looking for dipolar asymmetries in each stack. Assessing this method's feasibility may require a full mock observation, since it would sensitively depend on the galaxy sample, and the redshift-space dispersion within the filament. It would be important to rule out any mechanisms that might mimic a dipolar asymmetry in galaxy positions, such as filaments tilted even in real space. This would be a substantial investigation, which we leave to future work.

Going beyond galaxies, any ionized gas around the filament, likely co-rotating with the dark matter and galaxies, should scatter photons from the cosmic microwave background (CMB) in opposite directions in the two sides, by the kinematic Sunyaev–Zeldovich (kSZ) effect.
This would cause a dipole extending along the filament axes, imprinted on the CMB temperature map. It might be measured in a manner similar to how the kSZ effect has been detected around galaxies and galaxy clusters \citep{planckcollaboration/etal:2016,bocquet/etal:2019}. Again, assessing this detection's feasibility would be a substantial study, depending sensitively on both the amount of ionized gas rotation, and observational details, so we leave this to future work.
 
Even purely theoretically, it will be interesting to study how gas rotates within filaments. As in haloes, the collisionless dark matter likely has substantial radial velocities in a filament; a typical position near the axis would have many streams going both in and out. Gas, on the other hand, cannot multistream, and might be more rotationally supported. Gas is also subject to shocks and galactic feedback, adding stochasticity, but also can tend to be smoother than the dark matter \citep{harford/hamilton:2011}.

An exotic implication of our result is in a superfluid Bose-Einstein condensate dark-matter scenario, with sufficient self-interactions that spin-driven dark-matter vortices (associated with spin on the scale of the whole halo) might form. Wave, or fuzzy, dark matter ($\psi$DM) \citep[e.g.,][]{Hui2021} is similar to this scenario, except that without strong self-interactions, all that can realistically form in $\psi$DM are small vortex loops of scale the de Broglie wavelength \citep{HuiEtAl2021,SchobesbergerEtal2021}. But an additional, topological obstacle for spin-driven vortices to form in a halo is that vortices cannot end in a point, but arise as loops. Any vortex associated with a halo's spin would have to complete the loop outside the halo. This becomes more plausible (but still seems far-fetched) if it can thread the cosmic web through connected filaments, associated with their spin.

Filamentary gas since reionization is typically ionized, so if any processes might act differently on electrons and protons in filaments, e.g.\ in a Biermann battery mechanism \citep{kulsrud/etal:1997,gnedin/etal:2000,naoz/narayan:2013}, we speculate that rotation would generate a coherent magnetic field along the filament axis. A model with rotation in addition to shocks could help to understand the origin of cosmological seed magnetic fields. Also, an observational indication of a rotating filament could be a coherent magnetic field aligned with it, probed through e.g.\ synchrotron emission or Faraday rotation of a background polarized source \citep{brunetti/vazza:2020}.

Our finding is consistent with the standard tidal-torque theory (TTT) of the origin and evolution of angular momentum in large-scale structure. Net angular momentum arises from velocities on the outskirts of collapsing structures, and gets transported to the centre by gravity. The net rotational velocity around a filament is comparable to the known quadrupolar pattern, revising the existing picture. Originating from the same large-scale environment, filaments and their nearby galaxies/haloes are expected to share a similar large-scale spin field, and this provides another reason for why the spin of galaxies might be correlated with that of their nearby filaments.

We are unaware of a reason to think that our qualitative results, arising from gravity and the TTT, might differ in any near-concordance $\Lambda$CDM cosmology; indeed, our Millennium and Illustris results are similar. But quantitatively, filament spins likely do depend somewhat on cosmological parameters. In the TTT, angular momentum grows with the linear-theory velocity field and the Universe's expansion, $\sim D^{3/2}$ (with $D$ the growth factor) until `collapse,' after which its angular momentum is conserved \citep{white:1984}. This holds also in 2D, i.e.\ for infinitely long filaments \citep{neyrinck/etal:2020}. Consider an ensemble of filaments, for which all properties are held constant except for the growth of velocities before filaments collapse. Their degree of spin would depend on the amplitude of the velocity field at collapse, i.e.\ on parameters like $\Omega_{\rm M}$ and $\sigma_8$. But the cosmic web and its filaments would also change with cosmological parameters, so the picture is likely not as simple as that, and worth further study.

Our results indicate that the longest substantially rotating objects in the Universe are likely filaments. As we elaborate in \S\ref{sec:object}, candidate `objects' for this title include cosmic-web components: dynamically-defined haloes, filaments, walls, and voids. Haloes, including the largest galaxy clusters, rotate, but filaments can be much longer. Walls and voids have similar lengths as filaments, and even substantial widths or depths. However, walls, and especially voids, are unlikely to rotate as substantially and coherently as the rotating filaments.

\section*{Acknowledgments}
We acknowledge stimulating discussions with John Peacock, Catherine Heymans, Andy Taylor, Cheng Si, David Essex, and Beth Biller. We also thank an anomymous referee for helpful suggestions. 
Part of this work used the DiRAC@Durham facility managed by the Institute for Computational Cosmology on behalf of the STFC DiRAC HPC Facility (www.dirac.ac.uk). The equipment was funded by BEIS capital funding via STFC capital grants ST/K00042X/1, ST/P002293/1 and ST/R002371/1, Durham University and STFC operations grant ST/R000832/1. DiRAC is part of the UK National e-Infrastructure.
Late stages of this work were performed at the Aspen Center for Physics, which is supported by National Science Foundation grant PHY-1607611.
\textbf{Funding:} QX acknowledges support from the European Research Council under grant number 647112. 
MCN is grateful for funding from Basque Government grant IT956-16.
YC acknowledges the support of the Royal Society through the award of a University Research Fellowship and an Enhancement Award.
MAAC acknowledges support from Mexican grant DGAPA-PAPIIT IA104818.

\section*{Author Contributions}
QX performed almost all measurements and analysis, contributing to writing and ideas for methodology. MCN and YC conceived of the project, and led the analysis and writing. MAAC contributed some simulation data, ideas and analysis, and the animation in Fig.\ \ref{fig:rotatingfilament-frame}.

\section*{Data Availability Statement}
We will provide data and codes used for this paper upon reasonable request.

\bibliographystyle{mnras}

\bsp	\label{lastpage}
\end{document}